\documentclass{mn2e}
\usepackage{color}
\usepackage{graphicx}
\usepackage{dcolumn}
\usepackage{bm}
\usepackage{lscape}
\usepackage{epsfig}
\input{psfig}

\title[The $M_{\rm bh}$-$\sigma$ diagram] 
{An expanded $M_{\rm bh}$-$\sigma$ diagram, and a new 
calibration of active galactic nuclei masses}

\author[Graham et al.]
{Alister W.\ Graham$^1$\thanks{AGraham@astro.swin.edu.au}, 
Christopher A.\ Onken$^2$, 
E.\ Athanassoula$^3$ and 
F.\ Combes$^4$
\\
$^1$ Centre for Astrophysics and Supercomputing, Swinburne University
of Technology, Hawthorn, Victoria 3122, Australia.\\
$^2$ Mount Stromlo Observatory, The Australian National
 University, Private Bag, Weston Creek PO, ACT 2611, Australia.\\
$^3$ Laboratoire d'Astrophysique de Marseille (LAM), UMR6110,
CNRS/Universit\'e de Provence, Technop\^ole de Marseille Etoile, \\
38 rue Fr\'ed\'eric Joliot Curie, 13388 Marseille C\'edex 20, France.\\ 
$^4$ Observatoire de Paris, LERMA, 61 Av. de l'Observatoire, 75014 Paris, France.
}

\begin{document}
\label{firstpage}
\maketitle

\begin{abstract}

We present an updated and improved $M_{\rm bh}$-$\sigma$ diagram containing
64 galaxies for which $M_{\rm bh}$ measurements (not just upper limits) are
available.
Due to new and increased black hole masses at the high-mass end, 
and a better representation of barred galaxies at the low-mass end, 
the ``classical'' (all morphological type) $M_{\rm bh}$-$\sigma$ relation 
for predicting black hole masses is 
$\log(M_{\rm bh}/M_{\odot}) = 
(8.13\pm0.05) + (5.13\pm0.34)\log [\sigma/200\, {\rm km\, s}^{-1}]$,
with an r.m.s.\ scatter of 0.43 dex. 
Modifying the regression analysis to correct for a hitherto 
over-looked sample bias in which black holes with masses $<10^6 M_{\odot}$ are
not (yet) detectable, the relation steepens further to give $\log(M_{\rm bh}/M_{\odot}) =
(8.15\pm0.06) + (5.95\pm0.44)\log [\sigma/200\, {\rm km\, s}^{-1}]$.
We have also updated the ``barless'' and ``elliptical-only'' 
$M_{\rm bh}$-$\sigma$ relations introduced by Graham and Hu in 2008
due to the offset nature of barred galaxies.  These 
relations have a total scatter as low as 0.34 dex and currently define the
upper envelope of points in the $M_{\rm bh}$-$\sigma$ diagram.  They also have
a slope consistent with a value 5, in agreement with the prediction by Silk 
\& Rees based on feedback from massive black holes in bulges built by
monolithic-collapse.

Using updated virial products and velocity dispersions from 28 active galactic
nuclei, we determine that the optimal scaling factor $f$ --- which brings
their virial products in line with the 64 directly measured black hole masses 
--- is $2.8^{+0.7}_{-0.5}$.  This is roughly half the value reported by Onken et al.\
and Woo et al., and consequently halves the mass estimates of most 
high-redshift quasars. 
Given that barred galaxies are, on average, located $\sim$0.5 dex below the
``barless'' and ``elliptical-only'' $M_{\rm bh}$-$\sigma$ relations, we have
explored the results after separating the samples into barred and non-barred
galaxies, and we have also developed a preliminary corrective term to the velocity
dispersion based on bar dynamics. 
%
In addition, given the recently recognised coexistence of massive black holes and
nuclear star clusters, we present the first ever $(M_{\rm bh}+M_{\rm
nc})$-$\sigma$ diagram and begin to explore how galaxies shift from their
former location in the $M_{\rm bh}$-$\sigma$ diagram. 

\end{abstract}


\section{Introduction}

The $M_{\rm bh}$-$\sigma$ relation (Ferrarese \& Merritt 2000;
Gebhardt et al.\ 2000a) is important because: (i) it enables one to
predict supermassive black hole (SMBH) masses, $M_{\rm bh}$, in
galaxies for which only the bulge velocity dispersion, $\sigma$, is
known; (ii) it allows one to calibrate other relations which can then be
used to predict SMBH masses in active galactic nuclei (AGN) for which
$\sigma$ can not be readily measured; and (iii) it points toward a
physical connection between the nuclei of galaxies and the properties of
their host bulge.  SMBH masses themselves, plus their demographics, accretion
and activity is important for a number of reasons, in particular the influence 
that SMBHs are thought to have in dictating the growth of galaxies and galaxy clusters. 

Most AGN are too distant to spatially resolve 
any material which is predominantly under the 
dynamical influence of their black hole. 
Therefore, less direct methods to determine the masses of these black
holes are required.  Reverberation mapping (RM: e.g.\ Bahcall,
Kozlovsky \& Salpeter 1972; Blandford \& McKee 1982; Netzer \&
Peterson 1997) is the name given to observations which measure the
time delay between direct continuum emission from a central AGN and
the echoed emission-line signal from gas clouds in the so-called broad
line region (BLR: Seyfert 1943) surrounding the AGN (Shields 1974, see also 
Souffrin 1968). 
Given the constant speed of light, such time delays correspond to a distance,
$r$.  Coupled with the Doppler-broadened width $\Delta V$ of the
emission lines from the clouds, and assuming that their motion is
virialised
and dominated by the central black hole's gravity (e.g.\ Gaskell 1988, 2009a;
Koratkar \& Gaskell 1991; Wandel, Peterson \& Malkan 1999; Onken \& Peterson
2002), one can compute the virial product VP = $r\Delta V^2/G$.  To convert
these VPs into black hole masses requires the multiplication by a scaling
factor $f$ which is related to the geometry and orientation of the clouds and
effectively converts the measured velocity widths into an intrinsic Keplerian
velocity (Peterson \& Wandel 2000; Onken et al.\ 2004).

\begin{figure}
\includegraphics[angle=270,scale=0.56]{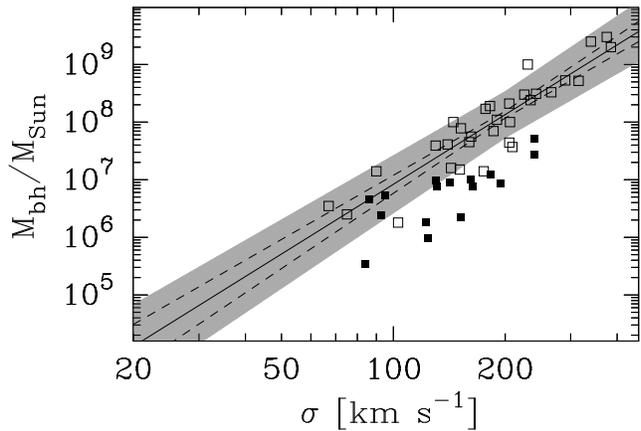}
\caption{ Open squares represent 31 predominantly inactive galaxies
  with direct SMBH mass measurements as tabulated by Tremaine et
  al.\ (2002).  The solid (and dashed) line is the relation (and associated
  1$\sigma$ uncertainty) reported by Tremaine et al.\ (2002).  The shaded
  region expands this domain vertically by 0.34 dex --- the 
  r.m.s.\ scatter about the relation.
  The 16 filled squares represent AGN with 
  (reverberation mapping)-derived virial products from Onken et
  al.\ (2004, their Table~3).  While in this figure an $f$-factor of 1
  has been used to plot these latter points, Onken et al.\ determined
  that an $f$-factor of $5.5\pm1.7$ was required to convert/elevate these
  AGN's virial products into black hole masses that agreed with the
  inactive galaxy sample in this diagram (see also Woo et al.\ 2010 who report
  $f=5.2\pm1.2$).  }
\label{Fig1}
\end{figure}

In diagrams that plot directly measured, and thus hopefully reliable, SMBH
masses $M_{\rm bh}$ obtained from nearby, predominantly inactive galaxies
versus some other galaxy property such as velocity dispersion, an empirical
calibration of the above $f$-factor can be performed by finding the value of
$f$ which yields the optimal overlapping agreement between the virial products
and the directly measured black hole masses.
Adopting a fixed value of $f=3$ (Netzer 1990), this calibration of RM masses
was first explored by Gebhardt et al.\ (2000b) and Ferrarese et al.\ (2001)
using stellar velocity dispersions for seven and six AGNs,
respectively.\footnote{The frequently used value of 3 is derived by taking the
isotropic assumption of $\sigma_{\rm 3D} = \sqrt{3}\sigma_{\rm 1D}$. However,
because Gebhardt et al.\ (2000b) and Ferrarese et al.\ (2001) made use of the 
emission line's full width 
at half maximum (FWHM), which Netzer took as 2$\sigma_{\rm 1D}$, they quote an
$f$-factor of 3/4 rather than 3. The RM results which we discuss adopt the second
moment of the line profile ($\sigma_{\rm line}$) as the measure of the
emission line width, and the $f$-factors we describe are appropriate to such
data.} 
Figure~\ref{Fig1} presents the results from Onken et al.'s (2004) first
ever empirical calibration of $f$, which assumed that the 16 local AGNs they
studied lie on the inactive galaxy $M_{\rm bh}-\sigma$ relation. 
The initial $M_{\rm bh}$-$\sigma$ relations for inactive galaxies 
possessed a small {\it total} root mean square (r.m.s.) scatter in the $\log M_{\rm bh}$
direction of $\sim$0.34 dex (Merritt \& Ferrarese 2001a; Tremaine et al.\
2002) and therefore appeared well-suited for the above task of calibrating the
$f$-factor. 
However, as more (barred) galaxies have been added to the $M_{\rm bh}$-$\sigma$ 
diagram, the r.m.s.\ scatter has increased above 0.4 dex, particularly at the low mass end 
(e.g.\ Graham 2008b; Gaskell 2009c). 
Indeed, Graham (2007a, 2008a), Hu (2008) 
and Graham \& Li (2009) have revealed that excluding barred
galaxies, or using 
only elliptical galaxies, results in the recovery of a tight barless, or
elliptical-only, $M_{\rm bh}$-$\sigma$ relation, 
a result reiterated by G\"ultekin et al.\ (2009b).  
The reason for this is not yet clear, but may be partly due to elongated 
orbits in bars or because bars can puff up the (face-on) central velocity
dispersion as they evolve (see Gadotti \& de Souza 2005; Gadotti \& Kauffmann
2009; Perez et al.\ 2009; Saha, Tseng \& Taam 2010).

Although bars are thin when they first form, vertical resonances (Combes \&
Sanders 1981; Combes et al.\ 1990; Raha et al.\ 1991) and possibly also hose instabilities
(Merritt \& Sellwood 1994) result in their growth out of the disc plane
leading to boxy-peanut-shaped bulges 
(Combes \& Sanders 1981; see also Illingworth 1981
and Kormendy \& Illingworth 1982). Similarly, the torques due to the bar
push gas inwards and can result in the formation of pseudo-bulges
(Kormendy \& Kennicutt 2004, Athanassoula 2005).
Athanassoula \& Misiriotis (2002, their figure~13) have shown how the
velocity dispersion can increase dramatically, both out of the disc plane and
along the length of the bar.
Therefore, barred galaxies have a mechanism by which they can migrate off the
$M_{\rm bh}$-$\sigma$ relation defined by non-barred galaxies.  Failing to
account for this offset population can not only bias one's estimates of SMBH
masses in AGN, but also impact on various evolutionary studies which may be
using galaxy samples with varying barred galaxy fractions at different
redshifts.

In Section~2.1 we introduce the (predominantly quiescent) galaxy
sample for which 64 direct supermassive black hole mass measurements
and velocity dispersions are available.  These galaxies are used to
define updated $M_{\rm bh}$-$\sigma$ relations.  In Section~2.2 we
introduce a sample of 28 AGN with available velocity dispersions and
explain how their RM measurements and line widths have been converted
into virial products.
In Section 3 we derive the optimal $f$-factor which brings the AGN in
line with the distribution of ($M_{\rm bh}, \sigma$) data for the
galaxies with direct SMBH mass measurements.
This factor provides the (sample average) calibration
needed to convert virial products into SMBH masses. 
We also identify a potential sample selection bias which, if uncorrected, results in 
underestimation of the slope of the $M_{\rm bh}$-$\sigma$ relation. 
A lengthy discussion is provided in Section~4, addressing the slope and
scatter of the $M_{\rm bh}$-$\sigma$ relation, sources of uncertainty on
measures of $M_{\rm bh}$ and $\sigma$, and implications of the new $f$-factor
for AGN masses.  Furthermore, 
a new equation
providing a first order correction for the influence of bar dynamics on the
host galaxy velocity dispersion is presented.  We also probe how the
coexistence of SMBHs and nuclear star clusters may alter our understanding of
the $M_{\rm bh}$-$\sigma$ diagram/relation.  Our main conclusions are
summarised in Section~5.

\section{An updated galaxy data set, and the $M_{\rm bh}$-$\sigma$ diagram/relation}

\subsection{Inactive galaxies}\label{SecIn}

Graham (2008b) provided a catalogue of ($M_{\rm bh}, \sigma$) values
for 76 predominantly inactive galaxies having direct SMBH mass
measurements which were carefully adjusted according to their adopted
distance.  Table~1 from that paper (based in part on Hu's 2008
compilation and  Ferrarese \& Ford 2005) provides ``reliable''
entries for 50 galaxies, 36 of which are non-barred systems.
G\"ultekin et al.\ (2009b) increased this sample\footnote{It is
  worth noting that there are differences among the
  adopted distances, and thus the adopted SMBH masses, between Graham
  (2008b) and G\"ultekin et al.\ (2009b), an issue already addressed
  by Graham \& Driver (2007b, their Section~2.1).}  with the addition
of 4 galaxies (G\"ultekin et al.\ 2009a)\footnote{G\"ultekin et
  al.\ (2009b) also includes 17 galaxies, not tabulated in Graham
  (2008b), for which only upper limits to their SMBH masses are
  available.  An additional 105 upper limits can be found in Beifiori
  et al.\ (2009).}  and we include these here after adjusting their
masses to the (Hubble constant)-independent distances reported by
Tonry et al.\ (2001).
Here we have, however updated the distance moduli that were reported by
Tonry et al.\ (2001, their Table~1), and used in Graham (2008b), by decreasing
their values by 0.06 mag,
thereby reducing the associated galaxy distances by $\sim$3 per cent, and thus
reducing the SMBH masses by this same amount.  This small correction stems
from Blakeslee et al.'s (2002, their Section~4.6) recalibration of the surface
brightness fluctuation method using the final Cepheid distances given by
Freedman et al. (2001, with the metallicity correction).  

In addition to the following 8 galaxies (IC~2560, NGC: 2974; 3079;
3414; 4552; 4621; 5813; 5846) that were not included by G\"ultekin et
al.\ (2009b), we further expand our ($M_{\rm bh}, \sigma$) catalogue
from Graham (2008b, his Table~1) with the inclusion of another ten
galaxies (see Table~\ref{Tab1}).  The following four were previously
considered to have uncertain SMBH masses 
(NGC~1068, Lodato \& Bertin 2003; 
NGC~3393, Kondratko et al.\ 2008; 
Abell 1836-BCG and Abell 3565-BCG, Dalla Bont\`a et al.\ 2009)\footnote{The
  references provided explain why these points are now considered reliable.}
while the following six are new galaxies: 
NGC 253 (Rodr{\'{\i}}guez-Rico et al.\ 2006); 
NGC~524 and NGC~2549 (Krajnovi\'c et al.\ 2009); 
NGC~1316 (Fornax A: Nowak et al.\ 2008); 
NGC 3368 and NGC 3489: (Nowak et al.\ 2010);
This gives a total sample of 64 galaxies 
with reliable SMBH masses. 

\begin{figure}
\includegraphics[angle=270,scale=0.41]{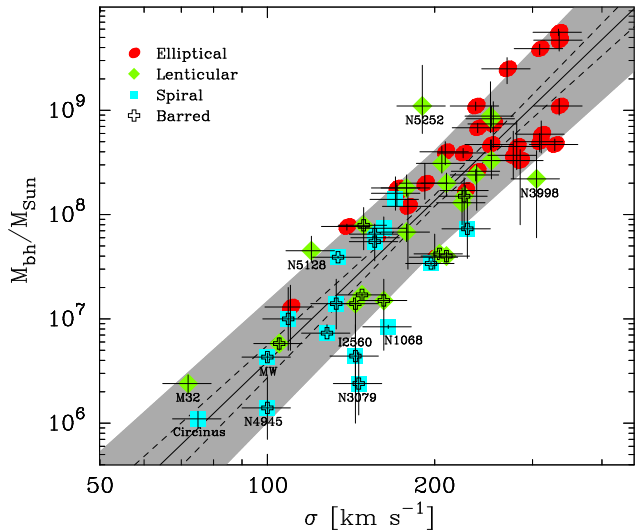}
\caption{ Updated $M_{\rm bh}$-$\sigma$ relation, containing 64 galaxies with
  reliable SMBH masses (see Table 1 from Graham 2008b, and Table 1
  from this paper).  The solid line is derived from a regression of 
$\log M_{\rm bh}$ on $\log \sigma$, assuming a 10 per cent uncertainty on the
  velocity dispersions $\sigma$ (see equation~\ref{Eq_full}).  
The dashed lines trace the 1-sigma 
uncertainty on this relation, while the shaded area extends this boundary by
  0.43 dex (the r.m.s.\ scatter about the relation) in the $\log M_{\rm bh}$
  direction.  The barred galaxies can be seen to dominate the distribution of points 
  below the best-fitting line at low-masses. 
}
\label{Fig2}
\end{figure}

Aside from this expansion from 50 to 64 galaxies, we have updated the mass for
the Milky Way's SMBH (Gillessen et al.\ 2009) and roughly doubled the SMBH
masses of NGC~3379, NGC~4486 and NGC~4649 (see Table~\ref{Tab1}). 
Recently published masses are increasingly secure due
to refinements such as the use of triaxial
orbit-based models rather than spherical or axisymmetric models, 
and better accounting for the range of orbital anisotropies and 
the influence of dark matter (Gebhardt \& Thomas 2009;
Shen \& Gebhardt 2009; van den Bosch \& de Zeeuw 2010).
Following G\"ultekin et al.'s (2009b) identification of an error in the SMBH
masses reported by Gebhardt et al.\ (2003), we have also increased the black hole 
masses, and their associated uncertainties, by a factor of 1.099 in the
following 9 galaxies:
%
NGC~821; NGC~2778; NGC~3384; NGC~3608; NGC~4291; 
NGC~4473; NGC~4564; NGC~4697; and NGC~5845. 
%
%
The expanded $M_{\rm bh}$-$\sigma$ diagram can bee seen in Figure~\ref{Fig2}.

\begin{table}
\caption{Expansion and update to Table~1 from Graham (2008b).}
\label{Tab1}
\begin{tabular}{lllll}
\hline
Gal.\ Id. & Type &  Dist.      & $\sigma$  &  $M_{\rm bh}$  \\
          &      &  Mpc        & km s$^{-1}$ &  $10^8 M_{\odot}$     \\
   1      &   2  &   3         &    4          &    5               \\
\hline
\multicolumn{5}{c}{New inclusions} \\
Abell 1836 & BCG &   157 [1a]  & 309 [6] & $39^{+4}_{-5}$ [6] \\
Abell 3565 & BCG &  40.7 [1a]  & 335 [7] & $11^{+2}_{-2}$ [6] \\
NGC 253   & SBc  &   3.5 [2]   & 109 [8]  & $0.1^{+0.1}_{-0.05}$ [11] \\ 
NGC 524   & S0   &  23.3       & 253      & $8.3^{+2.7}_{-1.3}$ [12] \\
NGC 1068  & Sb   &  15.2 [1b]  & 165 [9] & $0.084^{+0.003}_{-0.003}$ [13] \\ 
NGC 1316  & SB0  &  18.6 [3]   & 226      & $1.50^{+0.75}_{-0.80}$ [14] \\
NGC 2549  & SB0 [12] &  12.3   & 144      & $0.14^{+0.02}_{-0.13}$ [12] \\
NGC 3368  & SBab &  10.1       & 128      & $0.073^{+0.015}_{-0.015}$ [15] \\
NGC 3393  & SBab &  55.2 [1b]  & 197      & $0.34^{+0.02}_{-0.02}$ [16] \\ 
NGC 3489  & SB0  &  11.7       & 105      & $0.058^{+0.008}_{-0.008}$ [15] \\
NGC 3585  & S0   &  19.5       & 206      & $3.1^{+1.4}_{-0.6}$ [17] \\
NGC 3607  & S0   &  22.2       & 224      & $1.3^{+0.5}_{-0.5}$ [17] \\
NGC 4026  & S0   &  13.2       & 178      & $1.8^{+0.6}_{-0.3}$ [17] \\
NGC 5576  & E3   &  24.8       & 171      & $1.6^{+0.3}_{-0.4}$ [17] \\
\multicolumn{5}{c}{Updated data} \\
Milky Way & SBbc & 0.008 [4]  & 100 [10] & $0.043^{+0.004}_{-0.004}$ [4] \\
NGC 821   & E    &  23.4       & 200      &  $0.39^{+0.26}_{-0.09}$  [18] \\
NGC 2778  & SB0  &  22.3       & 162      &  $0.15^{+0.09}_{-0.10}$  [18] \\
NGC 3379  & E    &  10.3       & 209      &  $4.0^{+1.0}_{-1.0}$     [19] \\
NGC 3384  & SB0  &  11.3       & 148      &  $0.17^{+0.01}_{-0.02}$  [18] \\
NGC 3608  & E2   &  22.3       & 192      &  $2.0^{+1.1}_{-0.6}$     [18] \\
NGC 4291  & E2   &  25.5       & 285      &  $3.3^{+0.9}_{-2.5}$     [18] \\
NGC 4473  & E5   &  15.3       & 179      &  $1.2^{+0.4}_{-0.9}$     [18] \\
NGC 4486  & E0   &  15.6       & 334      &  $56^{+4}_{-4}$          [20] \\
NGC 4564  & S0   &  14.6       & 157      &  $0.60^{+0.03}_{-0.09}$  [18] \\
NGC 4649  & E1   &  16.4       & 335      &  $47^{+10}_{-10}$        [21] \\
NGC 4697  & E4   &  11.4       & 171      &  $1.8^{+0.2}_{-0.1}$     [18] \\
NGC 5128  & S0   &  3.8 [5]    & 120      &  $0.45^{+0.17}_{-0.10}$  [22] \\
NGC 5845  & E3   &  25.2       & 238      &  $2.6^{+0.4}_{-1.5}$     [18] \\
\hline
\end{tabular}

Unless otherwise specified, the distances have come from Tonry et al.\ (2001), 
after reducing their distance moduli by 0.06 mag (see Section~\ref{SecIn}). 
This small adjustment has been applied to all the galaxies from 
Graham (2008b, his Table~1) which used the Tonry et al.\ (2001) distance moduli. 
Unless otherwise specified, the velocity dispersions, $\sigma$, are the weighted values
from 
HyperLeda\footnote{http://leda.univ-lyon1.fr} (Paturel et al.\ 2003) as of March 2010.
The SMBH masses, $M_{\rm bh}$, have been adjusted to the distances given in column~3. \\
References: 
1a = NED: (Virgo + GA + Shapley)-corrected Hubble flow distance of the BCG's
host cluster; 
1b = NED: (Virgo + GA + Shapley)-corrected Hubble flow distance; 
2 = Rekola et al.\ (2005); 
3 = Madore et al.\ (1999); 
4 = Gillessen et al.\ (2009); 
5 = Karachentsev et al.\ (2007); 
6 = Dalla Bont\`a et al.\ (2009); 
7 = Smith et al.\ (2000); 
8 = Oliva et al.\ (1995); 
9  = Nelson \& Whittle (1995); 
10 = Merritt \& Ferrarese (2001a); 
11 = Rodr{\'{\i}}guez-Rico et al.\ (2006), a factor of 2 uncertainty has been
assigned here; 
12 = Krajnovi\'c et al.\ (2009); 
13 = Lodato \& Bertin (2003); 
14 = Nowak et al.\ (2008); 
15 = Nowak et al.\ (2010); 
16 = Kondratko et al.\ (2008); 
17 = G\"ultekin et al.\ (2009a); 
18 = Gebhardt et al.\ (2003), G\"ultekin et al.\ (2009b); 
19 = van den Bosch \& de Zeeuw (2010); 
20 = Gebhardt \& Thomas (2009); 
21 = Shen \& Gebhardt (2009); 
22 = Neumayer (2010). 
\end{table}









\subsubsection{The $M_{\rm bh}$-$\sigma$ relation(s)}

For predicting SMBH masses in other galaxies, one obviously desires a
relation with the minimal amount of scatter in the vertical ($\log
M_{\rm bh}$) direction.  This is achieved
with a non-symmetrical ordinary least squares regression of $\log
M_{\rm bh}$ on $\log \sigma$ (Feigelson \& Babu 1992).  
Using the BCES code from Akritas \& Bershady (1996) and assigning a 10 per
  cent uncertainty to the velocity dispersions
of the 64 galaxies with direct SMBH mass measurements, one obtains 
\begin{equation}
\log(M_{\rm bh}/M_{\odot}) =
(8.13\pm0.05) + (5.13\pm0.34)\log [\sigma/200\, {\rm km\, s}^{-1}], 
\label{Eq_full}
\end{equation}
with a total r.m.s.\ scatter of $\Delta = 0.43$ dex, 
and an intrinsic scatter of $\epsilon = 0.32^{+0.06}_{-0.04}$ dex, 
in the $\log M_{\rm bh}$ direction.  
This relation is shown in Figure~\ref{Fig2}. 
For reference, the equivalently produced relation using all 50 galaxies from 
Graham (2008b) has a slope and intercept of $4.87\pm0.36$ and $8.12\pm0.06$,
respectively, while Hu (2008) reported values of $4.59\pm0.32$ and $8.14\pm0.06$.
Using our updated data set, the slope drops to $4.86\pm0.31$, while the
intercept remains unchanged, when an uncertainty of 5 per cent --- the value
used by Tremaine et al.\ (2002) and G\"ultekin et al.\ (2009b) --- is assigned
to the velocity dispersions. 
Using Tremaine et al.'s (2002) modified FITEXY routine (Press et al.\ 1992)
yields consistent results, with a slope and intercept of 4.86 and 8.15,
respectively (when assigning a 5 per cent uncertainty to the velocity
dispersions).
Of the two commonly used uncertanties within the literature (namely, 5 and 10
per cent), we have elected to proceed using a 10 per cent
uncertainty on our velocity dispersions, as an inspection of HyperLeda
(Paturel et al.\ 2003, see also Section~6 in Nowak et al.\ 2010)
reveals that a 5 per cent uncertainty is probably overly optimistic.
Ideally though, one would like to have more accurate measurements and
measurement errors for the velocity dispersions (see Section~\ref{Sec_err}). 

When using (only) the 44 non-barred galaxies, and a 10 per cent uncertainty on
the velocity dispersions, the above regression of $\log M_{\rm bh}$ 
on $\log \sigma$ gives the relation 
\begin{equation}
\log(M_{\rm bh}/M_{\odot}) =
(8.25\pm0.06) + (4.57\pm0.35)\log [\sigma/200\, {\rm km\, s}^{-1}], 
\label{Eq_barless}
\end{equation}
with a total r.m.s.\ scatter of $\Delta = 0.37$ dex, 
and an intrinsic scatter of $\epsilon = 0.29^{+0.06}_{-0.05}$ dex, in the
$\log M_{\rm bh}$ direction.\footnote{When 
an uncertainty of 5 per cent is assigned to the velocity dispersion, the
slope of the ``barless'' $M_{\rm bh}$-$\sigma$ relation drops to $4.32\pm0.34$,
while the intercept basically remains the same.}
As already noted by Graham (2008a) and Hu (2008), 
a similar relation is obtained when using only the elliptical galaxies (see
Table~\ref{Tab2}).  The r.m.s.\ scatter of the elliptical-only 
$M_{\rm bh}$-$\sigma$ relation 
is 0.34 dex, notably less than the value of 0.43 dex for the 
standard (full sample) $M_{\rm bh}$-$\sigma$ relation. 

When using (only) the 20 barred galaxies, and a 10 per cent uncertainty on
the velocity dispersions, the regression of $\log M_{\rm bh}$
on $\log \sigma$ gives 
\begin{equation}
\log(M_{\rm bh}/M_{\odot}) =
(7.80\pm0.10) + (4.34\pm0.56)\log [\sigma/200\, {\rm km\, s}^{-1}],
\label{Eq_bar}
\end{equation}
with a total r.m.s.\ scatter of $\Delta = 0.36$ dex.  For convenience,
all of these $M_{\rm bh}$-$\sigma$ relations are tabulated in
Table~\ref{Tab2}.  This barred $M_{\rm bh}$-$\sigma$ relation is 0.45
dex (at $\sigma=200$ km s$^{-1}$) below the relation defined by
non-barred galaxies, and reiterates the offset nature of the two
populations first noted by Graham (2007a, 2008a,b) and Hu (2008).

\begin{table}
\caption{Assorted $\log (M_{\rm bh}/M_{\odot}) = \alpha + \beta \log (\sigma/200$ km s$^{-1})$ relations.}
\label{Tab2}
\begin{tabular}{@{}lcccc@{}}
\hline
Sample (size)    &     $\alpha$    &   $\beta$       &  $\Delta \log M_{\rm bh}$  &  $\epsilon_{\rm intrinsic}$ \\ 
                 &                 &                 &  [dex] &  [dex]  \\ 
\hline
\multicolumn{4}{c}{BCES regression of $\log M_{\rm bh}$ on $\log \sigma$} \\
Full (64)        &  $8.13\pm0.05$  &  $5.13\pm0.34$  &  0.43  &  $0.32^{+0.06}_{-0.04}$ \\
Barred (20)      &  $7.80\pm0.10$  &  $4.34\pm0.56$  &  0.36  &  $0.27^{+0.09}_{-0.07}$ \\
Non-barred (44)  &  $8.25\pm0.06$  &  $4.57\pm0.35$  &  0.37  &  $0.29^{+0.06}_{-0.05}$ \\ 
Elliptical (25)  &  $8.27\pm0.06$  &  $4.43\pm0.57$  &  0.34  &  $0.27^{+0.07}_{-0.05}$ \\
\multicolumn{4}{c}{BCES regression of $\log \sigma$ on $\log M_{\rm bh}$} \\
Full (64)        &  $8.15\pm0.06$  &  $5.95\pm0.44$  &  0.46  &  $0.35^{+0.06}_{-0.05}$ \\ 
Barred (20)      &  $7.94\pm0.18$  &  $5.40\pm1.22$  &  0.41  &  $0.31^{+0.06}_{-0.10}$ \\ 
Non-barred (44)  &  $8.24\pm0.06$  &  $5.32\pm0.49$  &  0.41  &  $0.30^{+0.07}_{-0.05}$ \\ 
Elliptical (25)  &  $8.22\pm0.09$  &  $5.30\pm0.77$  &  0.37  &  $0.29^{+0.08}_{-0.07}$ \\
\hline
\end{tabular}

Based on a 10 per cent uncertainty on the velocity
  dispersions $\sigma$.  The intrinsic dispersion $\epsilon_{\rm intrinsic}$ 
  pertains to the $\log M_{\rm bh}$ direction. 
\end{table}

Whilst we adopt the above approach for this section, we do note that Section~\ref{Sec_bias} 
points out, for the first time, a potential sample bias which, once corrected
for, results in steeper slopes for the various $M_{\rm
  bh}$-$\sigma$ relations.  These are given in the second half of Table~\ref{Tab2}. 
In Section~\ref{Sec_slope} we discuss the causes for the change in slope of the $M_{\rm
  bh}$-$\sigma$ relation.

\subsubsection{Predicting $M_{\rm bh}$}\label{Predict}

When predicting the SMBH mass of a new galaxy for which one knows the
velocity dispersion, the associated maximum 1-sigma uncertainty on the
black hole mass --- acquired by assuming uncorrelated errors on the velocity
dispersion $\sigma$ and both the slope and intercept of the $M_{\rm
  bh}$--$\sigma$ relation --- can be determined using Gaussian error
propagation.  For the linear equation $y=(b\pm\delta b)(x\pm \delta x)
+ (a\pm \delta a)$, one has an error on $y$ equal to
\begin{eqnarray}
\delta y & = & 
\sqrt{ (dy/db)^2(\delta b)^2 + (dy/da)^2(\delta a)^2 +
   (dy/dx)^2(\delta x)^2 } \nonumber \\
 & = & \sqrt{ x^2(\delta b)^2 + (\delta a)^2 + b^2(\delta x)^2 }. \nonumber
\end{eqnarray}
In the presence of intrinsic scatter in the $y$-direction, denoted by
$\epsilon$, the uncertainty on $y$ is
\begin{eqnarray}
\delta y = \sqrt{ x^2(\delta b)^2 + (\delta a)^2 + b^2(\delta x)^2 +
  \epsilon^2}. \nonumber
\end{eqnarray}
Using the standard $M_{\rm bh}$-$\sigma$ relation, given by 
equation~\ref{Eq_full}, we have that 
$x=\log(\sigma/200\, {\rm km\, s}^{-1})$, so $dx/d\sigma = 1/[\ln(10)\sigma]$, and therefore
\begin{eqnarray}
(\delta \log M_{\rm bh}/M_{\odot})^2 & = &
 [\log(\sigma/200\, {\rm km\, s}^{-1})]^2(0.34)^2 + (0.05)^2 \nonumber \\
 & & \hskip-40pt + [5.13/\ln(10)]^2 [\delta \sigma/\sigma]^2 + (0.32)^2.
\label{EqMerr}
\end{eqnarray}
This represents the uncertainty when predicting a new black hole mass 
from a velocity dispersion measurement $\sigma \pm \delta\sigma$.
Of course, if one knows the morphological type of the galaxy in question, then
the relevant equation from Table~\ref{Tab2} can be used to predict a
more accurate black hole mass.

Before proceeding, we note that several ways of distinguishing pseudo bulges
from classical bulges have been proposed in the literature. This includes
morphology, kinematic properties, the S\'ersic index and the distance from the
Kormendy (1977) relation.  Unfortunately there is no good agreement between
these methods and often authors using one method criticize the method(s) used
by others. On the other hand, whether a galaxy has a bar, or not, is a more
clear cut question. We thus believe that this is a much safer way of
addressing departures from the $M_{\rm bh}$--$\sigma$ relation.

\subsection{Active galaxies}


We have started with the homogenized time delays, $\tau$, and broad line
region velocity dispersion measurements for the 35 AGN listed in Table~6 of
Peterson et al.\ (2004).  Errors on both measurements are available, as are
multiple measurements for many galaxies.\footnote{Potential problems with
single epoch data are described in Gaskell (2009b).}  Below we 
describe how this data has been combined to acquire the single virial product,
\begin{equation}
c\tau\times\sigma_{\rm line}^2/G, 
\label{Eq_f}
\end{equation}
for each AGN.  Modulo the
yet-to-be-determined $f$-factor, this virial product represents the mass of each
AGN's central black hole. 

When multiplying uncorrelated numbers with errors, the relative error
on the product is the square root of the sum of the squares of the relative
errors in the individual numbers.  That is
\begin{eqnarray}
(x_i \pm \delta x_i)\times(x_j \pm \delta x_j)\times...(x_N \pm \delta x_N) 
  \nonumber \\
 = \left( \Pi_{i=1,N} x_i \right) \times \left( 1 \pm \sqrt{ \sum_{i=1,N}
\left(\frac{\delta x_i}{x_i}\right)^2} \right). \nonumber
\end{eqnarray}
When the numbers {\it are} correlated, the relative errors are simply added. 
Therefore, in deriving the virial product, $vp = c\tau_{\rm cent}
\sigma^2_{\rm line}/G$, from the 
$\tau_{\rm cent}$ and $\sigma_{\rm line}$ values tabulated by Peterson et
al.\ (2004), one has a relative uncertainty given by 
\begin{equation}
\frac{\delta vp}{vp} = \sqrt{ \left(\frac{\delta \tau}{\tau} \right)^2 + 
\left(\frac{\delta \sigma_{\rm line}}{\sigma_{\rm line}} +
\frac{\delta \sigma_{\rm line}}{\sigma_{\rm line}}\right)^2}. \nonumber
\end{equation}
These values are provided, for multiple measurements, 
in the final column of Table~6 from Peterson et al.\
(2004) and adopted here.

Assuming that these individual measurements, $vp_i \pm \delta vp_i$, of the
virial products are distributed normally about the true value $VP$, one can
use a maximum likelihood analysis to determine the (error-weighted average)
optimal value and its associated uncertainty.\footnote{The one simplification
is to average the slightly non-symmetrical errors associated with the virial
product measurements, providing the values of $\delta vp_i$ used above.}  From
the treatment of weighted averages by Taylor (1997, Chapter 7), one has that
\begin{equation}
VP = \frac{ \sum _{i=1,N} w_i vp_i}{ \sum_{i=1,N} w_i}, \, {\rm where} \,
w_i = \frac{1}{\delta vp_i^2}, \nonumber
\end{equation}
and the uncertainty on the value of VP is given by 
\begin{equation}
\delta VP = \frac{1}{\sqrt{\sum_{i=1,N} w_i}}. 
\label{Eq_dVP}
\end{equation}
The resulting (single) virial product for each galaxy is listed in
Table~\ref{Tab3} 
(c.f.\ Peterson et al.\ 2004, their Table~8; and Woo et al.\ 2010, which
appeared while we were preparing this work).  
Given that we ultimately find a different $f$-factor to Woo et
al.\ (2010), we felt that it was beneficial to show, in
Table~\ref{Tab3}, the exact data that we have used.

We have excluded PG~1211+143 and IC~4329A 
because the associated uncertainty on their virial 
products give values consistent with zero (Peterson et al.\ 2004).
We have also excluded the virial product for 3C 390.3 because it has a
double-peaked emission line profile.  
While NGC~5548 has the most 
extensive reverberation-mapping of all the AGN (Bentz et al.\ 2007),
frustratingly, it also has an irregular emission line profile. 
As noted by Zhu \& Zhang (2009) and Zhu, Zhang \& Tang (2009; see also Wandel,
Peterson \& Malkan 1999), the determination of the line width is problematic
for these last two galaxies.
To err on the side of caution, we 
present our analysis both with and without NGC 5548, finding consistent results.  
We have however included 12 new galaxies (see also Woo et al.\ 2010) 
for which velocity dispersions were 
not available at the time of Onken et al.'s (2004) analysis.  Table~\ref{Tab3}
lists 30 AGN with reverberation-mapping measures and velocity dispersions,
two of which we exclude, and 13 of which belong to barred galaxies (see Bentz et al.\
2009b, their Table~5).  These data are plotted in Figure~\ref{Fig3}.

\section{Analysis: The $f$-factor}

\begin{figure}
\includegraphics[angle=270,scale=0.82]{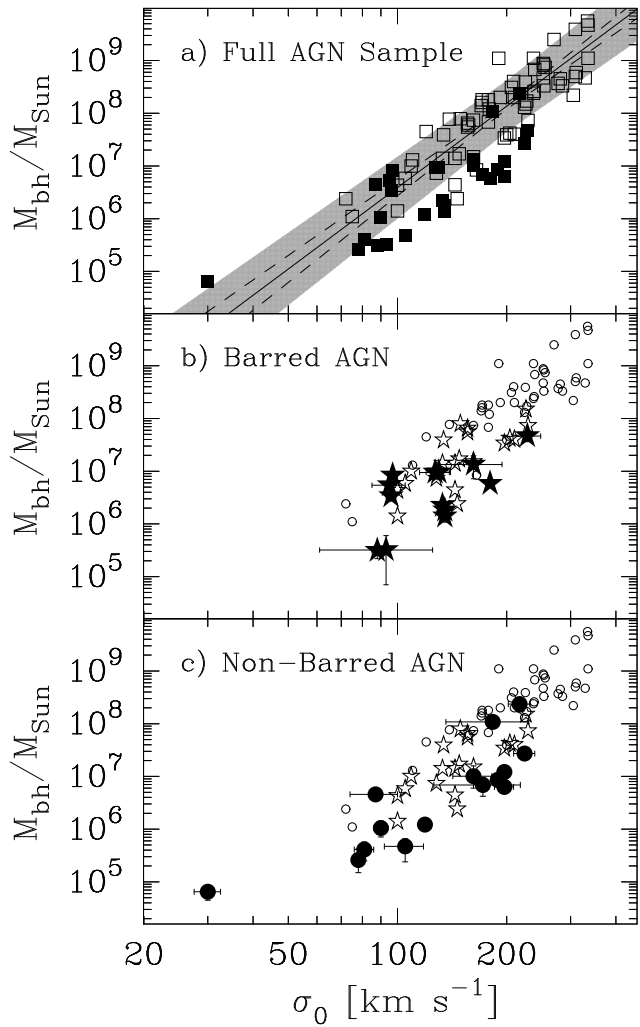}
\caption{
Panel a) An updated version of Figure~\ref{Fig1} using the galaxies described
in Section~2.  One can see that the full sample of 28 AGN (filled squares) with
(reverberation mapping)-derived virial products overlap with some of the 64 galaxies
with direct SMBH mass measurements (open squares).  The lines and shaded
region pertain to equation~\ref{Eq_full}.
In panels b) and c), both AGN and galaxies with bars are denoted by a star, while those without
by a circle.  Filled and open symbols pertain to our sample of AGN and
galaxies with direct SMBH mass measurements, respectively. 
In all panels, as in Figure~\ref{Fig1}, an $f$-factor of 1 has been used 
to plot the AGN virial products by the filled symbols. 
}
\label{Fig3}
\end{figure}

Figure~\ref{Fig3}a presents an updated version of Figure~\ref{Fig1}.  Again,
an $f$-factor of 1 has been applied to the AGN virial products.  Immediately
apparent is that the AGN virial products no longer appear quite so offset from
the directly measured SMBH masses, instead, the two populations somewhat
overlap.  The larger discrepancy seen in Figure~\ref{Fig1} appears to have been a result of
sample selection, specifically, due to the past under-representation of barred
galaxies with direct SMBH masses.

As noted previously, the $f$-factor is the normalization of the AGN virial product
which is used to estimate $M_{\rm bh}$ such that 
\begin{equation}
M_{\rm bh} = f \times \left( \frac{r\Delta V^2}{G} \right).
\label{eq_BH}
\end{equation} 
Here we seek the value of $f$ which best matches the AGN in the $M_{\rm
bh}$-$\sigma$ diagram with the $M_{\rm bh}$-$\sigma$ relation defined by the 
local sample of galaxies with direct SMBH mass measurements.

\begin{figure}
\includegraphics[angle=270,scale=0.82]{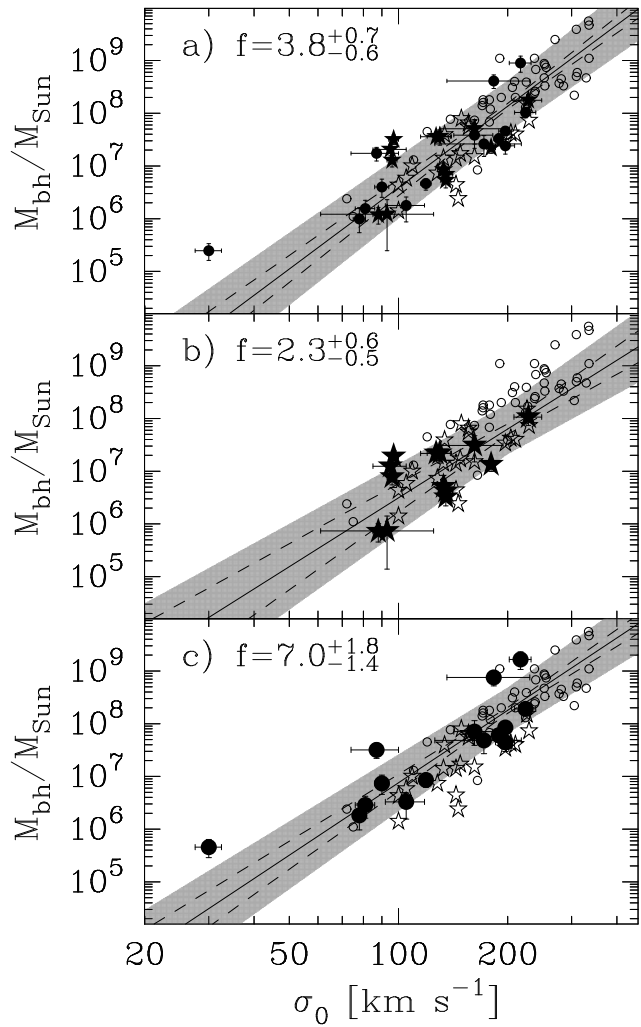}
\caption{
Similar to Figure~\ref{Fig3}, except that the optimal $f$-factor (inset in
figure) to bring the
AGN virial products in line with the corresponding $M_{\rm bh}$-$\sigma$ relation has been used. 
As in Figure~\ref{Fig3}, the filled symbols in panels a), b) and c) correspond
to the full, barred and non-barred AGN sample, while the open stars and circles
correspond to the barred and non-barred galaxies with direct SMBH mass
measurements. 
A linear regression of $\log M_{\rm bh}$ on $\log \sigma$ for the full, barred
and non-barred sample of galaxies with direct
SMBH mass measurements (see Section~\ref{SecIn}) has been used to construct
the three different relations (given by equations~\ref{Eq_full}, \ref{Eq_bar}
and \ref{Eq_barless}) shown in panel a), b) and c), respectively.
}
\label{Fig4}
\end{figure}

The $\chi^2$ value that we minimise to determine this optimal value of $f$ 
is given by the expression 
\begin{equation}
\chi^2 = \sum^N_{i=1} \frac{[\log(M_{{\rm bh},i}/M_{\odot}) - a -
  b\log(\sigma_i/200\, {\rm km\, s}^{-1})]^2}
{ [\frac{1}{\ln 10}\frac{\delta VP_i}{VP_i}]^2 + [\delta \log (M_{{\rm bh},i}/M_{\odot})]^2}, 
\end{equation}
where $a$ and $b$ are the intercept and slope of the $M_{\rm bh}$-$\sigma$
relation, $M_{{\rm bh},i}$ comes from equation~\ref{eq_BH}, 
$\delta \log(M_{\rm bh,i}/M_{\odot})$ is given in equation~\ref{EqMerr}, and 
$\delta VP_i$ is the uncertainty derived using equation~\ref{Eq_dVP} and listed 
in Table~\ref{Tab3}.  
Matching all 28 AGN to the standard (full sample) $M_{\rm bh}$-$\sigma$
relation given by equation~\ref{Eq_full} gives $f=3.8^{+0.7}_{-0.6}$.
This result is shown in Figure~\ref{Fig4}a.
Excluding NGC~5548, $f=3.6^{+0.7}_{-0.6}$.

As we have effectively seen, both here and in the literature,  
the inclusion of varying numbers of barred galaxies 
in the $M_{\rm bh}$-$\sigma$ diagram 
alters the best-fitting $M_{\rm bh}$-$\sigma$ relation that one obtains.  
We have therefore attempted to explore a derivation of the $f$-factor using 
only barred galaxies, and then again using only non-barred galaxies.

A visual inspection of the barred galaxies in Figure~\ref{Fig3}b and
\ref{Fig4}b reveals that the virial products from the 13 barred AGN overlap with
the territory occupied by the directly measured black hole masses of barred galaxies.
That is, an $f$-factor closer to a value of unity appears apt, and a formal
treatment produces a value of $f=2.3^{+0.6}_{-0.5}$.

Figure~\ref{Fig3}c reveals that 4 of the 15 unbarred AGN have
black hole masses which agree exceedingly well 
%
with the relation defined by the 44 non-barred quiescent galaxies when the AGN
virial products are multiplied by an $f$-factor of 1.  At the same time, 11 of
the AGN appear to require an $f$-factor significantly greater than a value of
$\sim$1-3.  Collectively, all 15 non-barred AGN generate an optimal
$f$-factor, relative to equation~\ref{Eq_barless}, of $=7.0^{+1.8}_{-1.4}$
($f=6.5^{+1.8}_{-1.4}$ when excluding NGC~5548), while using the 11 most
offset of these 15 AGN gives an unlikely $f$-factor in excess of 10.  Such a
high value of $f$ produces SMBH masses for four of the non-barred AGN which
are inconsistent with the upepr envelope of points in the $M_{\rm
  bh}$-$\sigma$ diagram defined using direct SMBH masses.

While none of these 11 galaxies display evidence of a bar, we do note that
three are peculiar in appearance (Ark~120, Slavcheva-Mihova \& Mihov 2010; 
Arp~151; and 3C~120), two may be
too distant or small to discern a bar (PG~2130+099 and Mrk~202, respectively), 
and another two (Mrk~279 and Mrk~590) were modelled with an ``inner
bulge'' by Bentz et al.\ (2009c), a feature that may be related to bars (Peng
et al.\ 2002). 
If we have over-looked the presence of bars in (some of) our allegedly
non-barred AGN sample, then our separation of AGN galaxies would obviously be
in error and we would thus be comparing the wrong types of galaxies in
Figure~\ref{Fig4}c, and likely over-estimating the value of $f$.
Unfortunately, we feel that it may
therefore be more appropriate at this time to prefer the $f$-value acquired
without any attempted division into barred and non-barred galaxies (i.e.\
Figure~\ref{Fig4}a).

Since commencing this project, several other potential mechanisms which may cause 
offset behavior in the $M_{\rm bh}$-$\sigma$ diagram have come to our
attention, all of which move galaxies below or rightward of the upper envelope
of points in the $M_{\rm bh}$-$\sigma$ diagram.  
First, as we discuss later in Section~\ref{Sec_NC}, the
offset nature of the barred galaxies in the $M_{\rm bh}$-$\sigma$ diagram may
also be, in part, due to the exclusion of the nuclear star cluster mass.  If 
so, this would further negate the appropriateness for the separation of 
galaxies solely on the basis of whether or not they contain a bar because 
(i) the mass of the neglected nuclear cluster may be important and 
(ii) non-barred galaxies can also be nucleated.
%
A second issue, briefly raised in section~\ref{Sec_RP}, pertains to
radiation pressure from the AGN.  This may cause the derived virial product to
erroneously fall below the $M_{\rm bh}$-$\sigma$ relation.
If the broad line region has a flattened spatial distribution, then inclined
AGN may appear to have narrower emission-line widths relative to those observed from an
edge-on orientation, and hence their black hole mass may be
under-estimated\footnote{AGN with BLRs
significantly more edge-on than average could actually end up with
over-estimated SMBH masses}.
In addition, if some AGN are effectively ignited by tidal interactions or
minor mergers driving gas inward (e.g., Dasyra et al.\ 2007), this external
trigger may potentially also elevate the velocity dispersion of the galaxy. 
%


Regrettably, the above three issues effectively handicap our ability to
proceed as we had hoped, and for the present time we fall back to the
standard approach used to date, which is to neglect the morphological
type in the analysis of the $f$-factor.  Nonetheless, our increased
sample size is more representative of the galaxy population at large,
and we feel that we are able to present a more appropriate measurement
of this calibration factor.


\subsection{Sample selection bias, and a re-derivation of the $f$-factor}\label{Sec_bias}


Collectively, we have not yet managed to directly, from spatially
resolved kinematics, 
measure the masses of SMBHs below about $10^6 M_{\odot}$ at the centres of
galaxies.  
This does not imply that such ``intermediate mass black holes'' 
(IMBHs) do not exist, indeed, evidence is accumulating that they probably 
do exist 
(e.g.\ Pox~52, Thornton et al.\ 2008; 
NGC~4395, Filippenko \& Ho 2003; 
Greene \& Ho 2004,2007; 
Dong et al.\ 2007; 
Naik et al.\ 2010; Seth et al.\ 2010).  We do however note that 
if some fraction of IMBHs form outside of galactic cores, or are kicked out 
by a gravitational radiation recoil event (e.g.\ Komossa \& Merritt 2008, and
references therein), then the relatively 
long dynamical friction time-scale for them to inspiral to the centre 
may result in some fraction of IMBHs still wondering outside of galaxy cores.  
Indeed, evidence for such a non-central IMBH may have already been discovered 
in galaxy ESO 243-49 (Soria et al.\ 2010; Wiersema et al.\ 2010). 

The non-detection of IMBHs, with resolved gravitational sphere's of influence, 
at the centres of galaxies 
suggests that a (typically neglected) selection bias may exist within the 
$M_{\rm bh}$-$\sigma$ diagram.  If our survey selection is such that SMBHs
with masses less than $\sim$$10^6 M_{\odot}$ are excluded, then it becomes
necessary to modify the type of linear regression which is used if one is to 
construct a non-biased $M_{\rm bh}$-$\sigma$ relation.  To avoid this 
sample selection bias requires a regression which minimises the residuals in
the $\log \sigma$ direction.  A discussion of this problem and solution can
be found in Lynden-Bell et al.\ (1988. their Figure~10).

\begin{figure*}
\includegraphics[angle=270,scale=0.86]{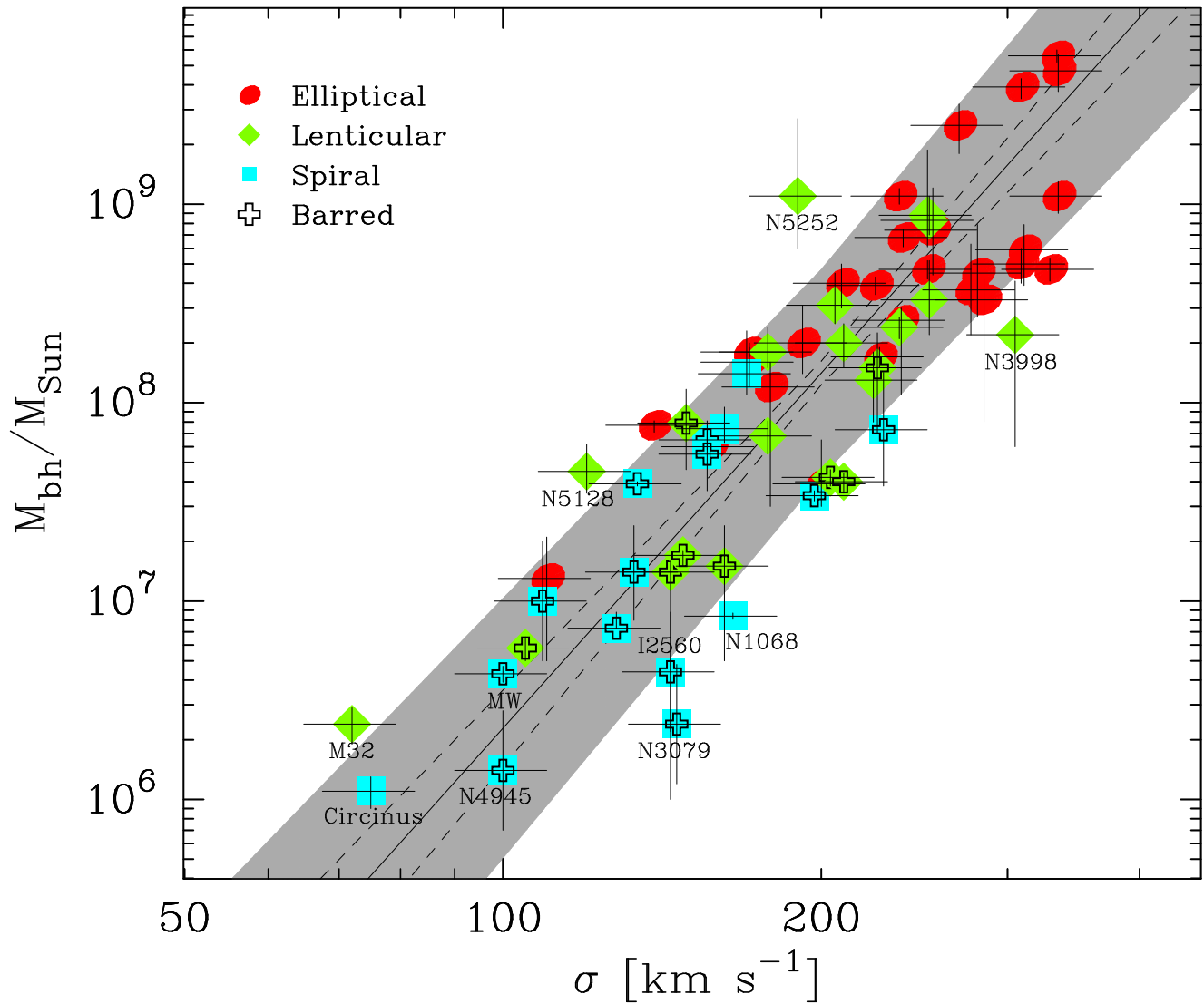}
\caption{
Similar to Figure~\ref{Fig2}, except that 
a regression of $\log \sigma$ on $\log M_{\rm bh}$ for the 64 galaxies with direct
SMBH mass measurements has been used (see Table~\ref{Tab2}). 
}
\label{Fig5}
\end{figure*}

\begin{figure}
\includegraphics[angle=270,scale=0.82]{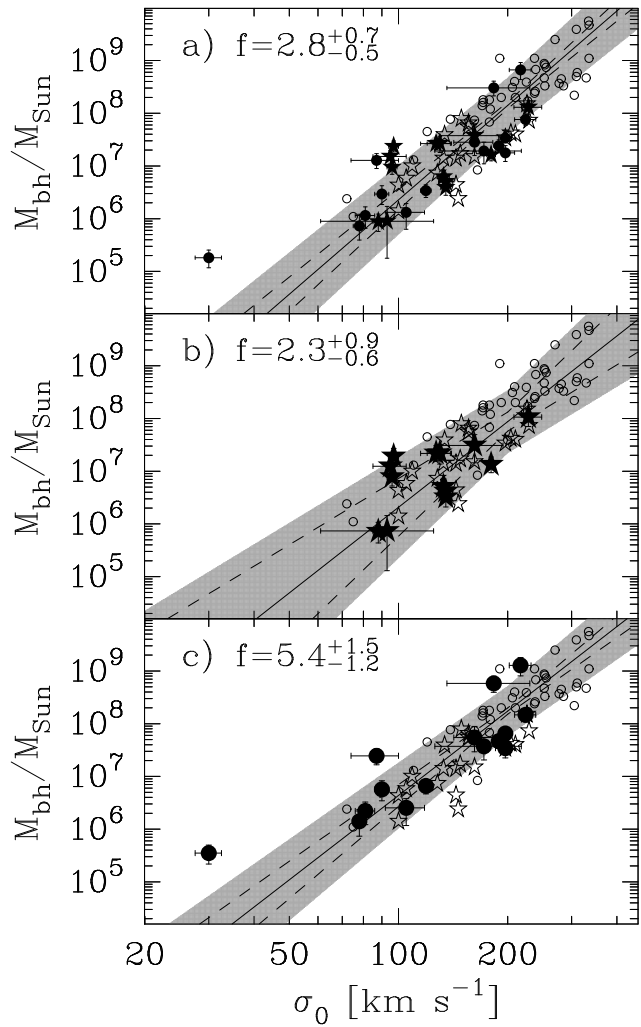}
\caption{
Similar to Figure~\ref{Fig4}, except that 
a regression of $\log \sigma$ on $\log M_{\rm bh}$ for the galaxies with direct
SMBH mass measurements has been used to construct 
the three linear regressions in each panel (see Table~\ref{Tab2}). 
Excluding NGC~5548 gives $f=2.6^{+0.6}_{-0.4}$, $2.3^{+0.9}_{-0.6}$ and
$4.9^{+1.4}_{1.1}$ in panels a), b) and c), respectively.
}
\label{Fig6}
\end{figure}

We have therefore repeated the previous analysis using an $M_{\rm
bh}$-$\sigma$ relation constructed from a linear regression of $\log \sigma$
on $\log M_{\rm bh}$ for (i) the full sample (see Figure~\ref{Fig5}), (ii) the
barred galaxies and (iii) the non-barred galaxies.  The resulting relations,
provided in Table~\ref{Tab2}, are steeper than previously
obtained.\footnote{The slope of the VP-$\sigma$ relation for the 28 AGN is
$5.09\pm0.88$, consistent with the associated slope of the $M_{\rm
bh}$-$\sigma$ relation ($5.95\pm0.44$) for the 64 predominantly inactive
galaxies.  We have therefore not explored changes in $f$ as a function of
black hole mass or velocity dispersion.}
Consequently, this results in a reduction to the optimal
f-factors which are shown in Figure~\ref{Fig6}.
Using the 28 AGN and 64 predominantly-inactive galaxies, the optimal "bias-free" 
$f$-factor is $2.8^{+0.7}_{-0.5}$. 
This value is a factor of 2 less than reported by Onken et 
al.\ (2004) and Woo et al.\ (2010), but in good agreement 
(perhaps for different reasons) with the value of 
$3.1^{+1.3}_{-1.5}$ from Marconi et al.\ (2008, see section~\ref{Sec_RP}).
Excluding the non-barred AGN NGC~5548 results in $f$-factors 
  of $f=2.6^{+0.6}_{-0.4}$, $2.3^{+0.9}_{-0.6}$ and $4.9^{+1.4}_{1.1}$ in
  Figures~\ref{Fig6}a), b) and c), respectively.

\section{Discussion}

\subsection{The slope, and scatter, of the $M_{\rm bh}$-$\sigma$ relations}\label{Sec_slope}

The slope of the $M_{\rm bh}$-$\sigma$ relation has received a lot of
attention since Gebhardt et al.\ (2000a) reported a value of $3.75\pm0.3$ while
Ferrarese \& Merritt (2000) reported a value of $4.8\pm0.5$.  Such slopes had 
been predicted because of the way feedback from
accretion onto the central massive black hole re-directs energy and momentum
back into a galaxy, establishing a correlation in which $M_{\rm bh}$ scales with
$\sigma^4$ (Fabian 1999) or $\sigma^5$ (Silk \& Rees 1998; Haehnelt, Natarajan
\& Rees 1998). 
%
However, 
disagreement over the exact slope has continued, with Tremaine et 
al.\ (2002) claiming a value of $4.01\pm0.32$ and Ferrarese \& Ford (2005)
advocating $4.86\pm0.43$.  An insightful discussion as to why those measured
slopes varied, depending on the type of regression used, 
is provided by Novak, Faber \& Dekel (2006), see also Feigelson \& Babu (1992)
for background understanding. 

Relative to the ``classical'' $M_{\rm bh}$-$\sigma$ relation based on the use
of galaxies of every morphological type, Graham (2007a, 2008a) and Hu (2008)
have revealed that there is a tighter relation based on the use of
elliptical-only galaxies, or non-barred or pseudo bulge galaxies.  In
addition, these new relations have a different, shallower, slope than the one
obtained using every galaxy.
Using a symmetrical regression analysis, 
G\"ultekin et al.\ (2009b) recently reported a slope of $4.21\pm0.45$ for 41
allegedly non-barred galaxies with available SMBH mass measurements, almost
identical to the slope of 4.28 previously reported by Graham (2008b) when
using a sample of 36 non-barred galaxies and the same 5 per cent uncertainty
assigned to the velocity dispersions.
%
Using a non-symmetrical regression of $\log M_{\rm bh}$ on $\log \sigma$, 
Graham (2008a) had also shown that the slope of the 
``elliptical-only'' $M_{\rm bh}$-$\sigma$ relation was as low as 
$3.68\pm0.25$.  Hu (2008) also constructed such an ``elliptical-only $M_{\rm
  bh}$-$\sigma$ relation'', finding a slope of $3.82\pm0.36$ when using a
5 and 10 per cent uncertainty for the early- and late-type galaxies,
respectively. 
Using an uncertainty of 5 per cent for the velocity dispersions, 
G\"ultekin et al.\ (2009b) subsequently reported a slope of 3.96 for their
``elliptical-only'' $M_{\rm bh}$-$\sigma$ relation, and a slope of 
3.86 when using their early-type galaxies. 
%


We have shown, with an updated and expanded data set, that the above slopes
have increased. The reason is in part due to new data and the doubling of
some previous SMBH masses at the high-mass end.  Table~\ref{Tab2} reveals 
that a symmetrical 
treatment of the data --- obtained by averaging the slopes obtained when
regressing $\log M_{\rm bh}$ on $\log \sigma$ and $\log \sigma$ on $\log
M_{\rm bh}$ --- 
yields a slope of (4.57+5.32=) 4.95 and
(4.43+5.30=) 4.87 for the non-barred and elliptical-only galaxies,
respectively.  
While this is equivalent to the (symmetrical regression)-derived 
slope reported by Ferrarese \& Merritt (2000)
and Ferrarese \& Ford (2005), one needs to keep in mind that their slope
pertained to a classical (barred plus unbarred) galaxy sample.  
Curiously, the optimal value for the slope of the barless and
elliptical-only $M_{\rm bh}$-$\sigma$ relations 
has changed from $\sim$4 to $\sim$5, consistent with the prediction by Silk \& Rees (1998)
based on feedback from SMBHs in bulges built by monolithic-collapse. 
Given the potential sample bias which currently excludes SMBH masses less than
about one million solar masses, the relations in the lower half of
Table~\ref{Tab2} should be preferred; these are consistent with a slope of 5. 
The slope for the full galaxy sample (barred plus unbarred galaxies) is 
steeper still. 
The reason for this is also partly due to the increase of some 
SMBH masses at the high-mass end, and the inclusion of more barred
galaxies at the low-mass end.  
Given that the barred galaxies appear to define their own offset relation
relative to the non-barred galaxies, coupled with the observation that they 
have velocity dispersions which only span the lower-half of the range spanned 
by the non-barred galaxies, their increased numbers results in a
steepening of the ``classical'' $M_{\rm bh}$-$\sigma$ relation. 
Using all 64 galaxies yields a slope between 5 and 6 
(see Table~\ref{Tab2}), a result which appears to be subject to the relative 
numbers of differing morphological type (barred versus unbarred galaxies) that
one includes.  Although, this latter remark may need to be revoked depending 
on the role that nuclear star clusters play. 

Finally, it is worth noting that 
the scatter about the ``classical'' $M_{\rm bh}$-$\sigma$ relation remains in excess of
0.4 dex.  It would thus appear that the velocity dispersion is not the 
fundamental/sole parameter driving the host galaxy connection with the
SMBHs.\footnote{Technically, it is the intrinsic scatter rather than the total
  scatter which reveals what physical quantity may control the SMBH mass.
  However, uncertainty in the measurement errors of the velocity dispersion
  render the intrinsic scatter a rather unreliable quantity.  Nonetheless, using a 5
  per cent uncertainty, G\"ultekin et al.\ (2009b) report an intrinsic scatter
  of $0.44\pm0.06$ dex for their full sample $M_{\rm bh}$-$\sigma$ relation.}
The $M_{\rm bh}$-$n$ relation (Graham \& Driver 2007a), which uses the {\it
  major-axis} S\'ersic index $n$, has a total scatter of only 0.31 dex. 
In addition, 
(Graham 2007b) has shown that the 
$M_{\rm bh}$-$L$ relation (McLure \& Dunlop 2002; Erwin et al.\ 2003;
Marconi \& Hunt 2003) currently has a total scatter of only 0.33 dex 
when (i) based on 
near-IR CCD images rather than optical photographic plates, (ii) using 
$R^{1/n}$ rather than $R^{1/4}$ modelling of the bulge light, (iii) 
after the correct identification of disc 
galaxies and thus the correct separation of bulge and disc light,
and (iv) after applying internal dust corrections to spiral and lenticular 
galaxies. 
When such corrections are made, rather than
using data from de Vaucouleurs et al.\ (1991), as done by 
G\"ultekin et al.\ (2009b), 
the uncertainty on the slope and intercept of the $M_{\rm bh}$-$L$ relation 
are reduced, and the 
{\it intrinsic} scatter drops from 0.38 dex to 0.30 dex. 
%
Consequently, one may be tempted to conclude that the bulge 
luminosity and major-axis S\'ersic index 
appear to be capable of predicting black hole masses with more
accuracy than the stellar velocity dispersion.  However, we note that
Graham \& Driver (2007a) contains only six barred galaxies, while
Graham's (2007b) re-analysis of the Marconi \& Hunt (2003) galaxy
sample contains only four barred galaxies.  If, after an accurate
modelling of the stellar light distribution, barred galaxies are not
found to be offset in say the $M_{\rm bh}$-$L_{\rm bulge}$ diagram, and
the scatter remains small, it would suggest that the $\sigma$ values may be
responsible for these galaxies' offset nature in the $M_{\rm
  bh}$-$\sigma$ diagram.  If, on the other hand, barred galaxies are
similarly offset in the $M_{\rm bh}$-$L$ diagram as they are in the
$M_{\rm bh}$-$\sigma$ diagram, it may be suggestive of an underweight
SMBH mass relative to the relation / upper-envelope defined thus far
(see Batcheldor 2010).

\subsection{Sources of uncertainty in $M_{\rm bh}$ and $\sigma$}
\label{Sec_err}

There are many sources of uncertainty in the values of $M_{\rm bh}$ and
$\sigma$.  While typically not spoken about, they are present in all the
previous studies of the $M_{\rm bh}$-$\sigma$ relation, and also this one. 
While Peterson (2010) have already provided cautionary remarks on the
reliability of reverberation-mapping derived SMBH masses, Cappellari et al.\
(2010) do the same for direct measurements of gas and stellar dynamics around SMBHs.
Attempting to resolve these issues is not only beyond the intended scope of
this paper, but would certainly make it quite unwieldy.  Nonetheless, we felt
it was appropriate, and hopefully helpful, to list a few of our own concerns.

\subsubsection{Cautionary remarks regarding $M_{\rm bh}$}

In regard to the SMBH mass, we list three issues.
As we have seen from Table~\ref{Tab1}, observers continue to refine/modify
their SMBH mass measurements as new data becomes available and new techniques
are implemented (e.g., Valluri, Merritt \& Emsellem 2004; Cappellari et al.\ 2010).  
Sometimes this reveals an under-estimation of past error bars on SMBH masses.  Point 
1) Using triaxial, rather than oblate or spherical, models 
can result in factor of two changes and possibly more (van den Bosch \&
de Zeeuw 2010).  
2) Failing to allow for a NC with a different stellar
mass-to-light ratio from the underlying bulge will bias one's mass measurement of
a SMBH. 
3) A proper dynamical treatment of bars in regard to the mass measurement of
SMBHs, has in general not been undertaken, and Krajnovi\'c et al.\ (2009) have remarked that 
``The bias introduced by modelling a likely barred galaxy using a model with a
static, axisymmetric potential has so far not been well explored.''

\subsubsection{Cautionary remarks regarding $\sigma$}
\label{Sec_Hot}

At the top end of the $M_{\rm bh}$-$\sigma$ relation, one is working with massive
elliptical galaxies which can have steep velocity dispersion gradients near 
their centre.  We are referring to the profile beyond the SMBH's sphere
of influence, where the velocity dispersion profile is steep due to 
the high central concentration of stars, as traced by the S\'ersic index
(e.g.\ Trujillo, Graham \& Caon 2001).  
Under poorer seeing conditions, the measured central velocity
dispersion will be under-estimated as one effectively samples more of the
surrounding light which has a lower velocity dispersion (cf. Graham et al.'s
1998 value of $353\pm19$ km s$^{-1}$ and D'Onofrio et al.'s 1995 value of
$420\pm27$ km s$^{-1}$ for NGC~1399).  
Furthermore, given the declining nature
of velocity
dispersion profiles with radius, especially in big elliptical galaxies, the
use of an average luminosity-weighted velocity dispersion over a larger
aperture should be systematically different from the central velocity
dispersion.  
The use of total, infinite aperture, luminosity-weighted velocity
dispersion, rather than central velocity dispersion, would result in an 
$M_{\rm bh}$-$\sigma_{\rm total}$ relation which is 
considerably steeper than the $M_{\rm bh}$-$\sigma_{\rm central}$ relation.
This is readily appreciated by looking at the aperture velocity dispersion
profiles in Graham \& Colless (1997, their figure~8) 
and Simonneau \& Prada (2004, their figure~5). 

At the lower-mass end of the $M_{\rm bh}$-$\sigma$ relation,
when dealing with disc galaxies, the velocity dispersion can be 
over-estimated due to strong rotational gradients within the inner region used
to measure $\sigma$.  The transition in rotational velocity ($V_{\rm rot}$), 
from say $+$220 km s$^{-1}$ to $-$220 km s$^{-1}$ (the rotational
velocity of our Sun, e.g.\ Liu \& Zhu 2010) 
across the centre of a disc galaxy can happen over 
a small radial range.  Given that some fraction of this rotational velocity change will
occur within one's adopted aperture or fraction of the spectrograph's
slit-length, the absorption line profile from the integrated flux can be
highly broadened in non-(face on) discs, thereby significantly increasing the measured velocity
dispersion above the true value (see Epinat et al.'s 2010 work with emission
lines).  Needless to say, given that the
apparently offset galaxies in the $M_{\rm bh}$-$\sigma$ relation tend to be
disc galaxies (Graham 2008a; Hu 2008; G\"ultekin et al.\ 2009b),
this scenario is rather interesting. 
In passing we remark, perhaps for the first time, 
that this issue may be a factor in explaining the 
high velocity dispersion measurements reported for some compact galaxies 
at high redshifts (van Dokkum, Kriek \& Franx 2009). 

Velocity dispersions can come from spectrograph slits placed along the major
axis, the minor axis, or the bar axis.  Apertures and slits can be of varying
radius and length (and thus sample varying fractions of a bulge's half-light radius); together
with different galaxy distances, plus varying bar lengths, this can all act to
modify the velocity dispersion that is actually measured.  Detailed
measurements, from integral field spectrographs, for the 64 galaxies with
direct SMBH mass measurements would be a welcome contribution (see Batcheldor
et al.\ 2005).  Such uniform data, sampling spectral lines that are not particularly
sensitive to young stars (such as the Ca Triplet lines), and whose acquisition
is admittedly beyond the scope of
this paper, would be preferable to the publicly available, hetereogeneous mix
of velocity dispersions which are, thankfully, homogenised by HyperLeda
(Paturel et al.\ 2003) following the precepts of McElroy (1995) 
to produce the galaxy velocity dispersions that we and 
others use.  Having a combination of data from many authors does, however, reduce
the impact that systematic biases --- which may be present in some individual observing
programs --- can have. 

The problem with young stars, alluded to above, is that they shine very
bright, and can bias, and even dominate, one's luminosity weighted velocity
dispersion (e.g.\ Wozniak et al.\ 2003).  One may therefore have a situation
where a cold disc has been able to form new stars, and this young nuclear disc
dominates the central emission but does not represent the dynamics of the
classical bulge.  Similarly, one may have a significant contribution from a
relatively cold nuclear or inner bar.  This may result in ``sigma drops''
where the central velocity dispersion is actually depressed (e.g.\ Franx et
al.\ 1989; D'Onofrio et al.\ 1995; Graham et al.\ 1998; H\'eraudeau \& Simien
1998) or the value at the ends of the inner bar is lowered (de
Lorenzo-C\'aceres et al.\ 2008).

Studying the nuclear region of three barred galaxies, Emsellem et al.\ (2001)
observed the near-infrared CO-band head lines, where dust obscuration is far
less of a concern than in the optical.  Their observed sigma-drops, of
generally $<20$ per cent, were due to nuclear bars or discs, visible in part
because of their sub-arcsecond seeing.  Chung \& Bureau (2004) subsequently
reported on the detection of sigma-drops in a larger sample of edge-on
galaxies. They too found decrements of up to 15-20 per cent due to likely
central discs.  These papers revealed that the sigma-drops are only present in
a small fraction of galaxies, and that they provide a smaller perturbation
compared to that capable of a large scale bar.


\subsection{A first order correction to $\sigma$ for bar dynamics}

As already discussed, we found that the 20 galaxies with large scale bars,
from among the 64 predominantly inactive galaxies with direct SMBH mass
measurements, are somewhat offset from the barless $M_{\rm bh}$-$\sigma$
relation towards higher velocity dispersions. In this section we discuss how
this may be linked to orbital bar dynamics. 
We do not, however, consider the potential bias due to nuclear/inner bars, or
nuclear discs with young stars which can exist in both barred and
non-barred galaxies.  In what follows we offer a simplistic, statistical 
correction for the large scale bars. 

When measuring the velocity dispersion in the innermost portion of an edge-on
SA galaxy, one has contributions mainly from the bulge orbits.  This is
because in the zero-th order approximation, the orbits of disc stars are
circular.  Consequently, any disc stars seen (in projection) at the very
center of the galaxy will have velocities perpendicular to our line-of-sight. 
Of course the disc star orbits are not exactly circular, and the measurements
refer not only to the very center but to a region around it (see the previous
subsection).  Inspite of this, the contribution of the disc to the velocity
dispersion will be small compared to that of the bulge and can be neglected.
This is not the case for SB galaxies.

The situation is more complex in a barred galaxy. Here the building blocks are
closed periodic orbits, whose shape changes a lot as a function of location
within or around the bar (e.g.\ Athanassoula 1992, her figure~3). If these are
stable, they trap around them regular orbits, which contribute to the velocity
dispersion, in the same way as the orbits trapped around the near-circular
orbits in SA galaxies. In the SB galaxies, however, there are further effects.
Since the orientation of the bar is random with respect to the line of nodes,
the velocity along the orbit is not necessarily perpendicular to the
line-of-sight unless the bar is along the line of nodes, or is perpendicular
to it (but, in this second case, for a much smaller inner radial range). 
Furthermore, some periodic orbits have loops which, if projected on the
centre, can further increase the velocity dispersion.  Readers wishing to
understand these effects better can find more extensive explanations in Bureau
\& Athanassoula (1999), where individual periodic orbits are analysed.  The
bottom line is that the orbital structure of the bar will increase the
observed velocity dispersion beyond that of the bulge alone, and will thus
partly explain the (positive) sign of the deviation of barred galaxies from
the regression line obtained from the remaining sample. Furthermore, the
simulations analysed in Bureau \& Athanassoula (2005) show that these effects
can be quite strong and may enhance the observed velocity dispersion by as
much as 10 -- 40 per cent, depending on the bar's strength and orientation
(Bureau \& Athanassoula 2005, their figure~1). This range is in agreement with
the amplitude of the observed deviations. It is thus clear that these
deviations may well be due to the orbital structure characteristics due to the
bar potential.


Here we provide a simplistic estimate of the affect from large-scale bars 
on the velocity dispersion (in the disc plane).  Imagine a bar with 
major- and minor-axis lengths of 4 and 2 kpc, respectively.  The bar thus has 
$b/a=0.5$ and is some 8 kpc long.  Seen end on, the bar is 4 kpc wide.  
To give some indication as to what fraction of the bar may be sampled by apertures
used to measure a galaxy's central velocity dispersion, 
1$\arcsec$ corresponds to $\sim$0.1 kpc at a distance of 20 Mpc. 
Central velocity dispersions are typically measured within the inner couple
of arcseconds, thus sampling the inner $\pm$0.1 kpc in this example. 
The main family of orbits sustaining the bar are x1 orbits, which are like ellipses 
that are aligned with the bar and centred on the bar; these are of course different
from purely radial orbits.  As noted above, 
when this bar is viewed end-on, and at 90 degrees 
from this orientation (i.e.\ viewing the full length of the bar), the central
stellar orbits will again be perpendicular to our line of sight. 
This, however, will not be the case at intermediate viewing angles. 
That is, the increase in velocity dispersion will be due to elliptical orbits
entering the 0.1-0.2 kpc aperture when the bar has a position angle different
from 0 or 90 degrees, with the maximum when the bar is at 45 degrees, contary
to what one may at first imagine.

To roughly correct the observed central velocity dispersion ($\sigma_{\rm
  obs}$) for broadening by non-circular disc motions due to a large-scale bar,
and thus acquire the bulge velocity dispersion ($\sigma_{\rm bulge}$), may
require the use of an expression like
\begin{equation}
\sigma^2_{\rm bulge} = \sigma^2_{\rm obs} -
 C\, V_{\rm rot}^2\, {\rm sin}^2(i)\times {\rm sin}^2(PA)\times {\rm
 cos}^2(PA),
\label{eq_Fran}
\end{equation} 
where $i$ is the inclination of the disc such that $i=0$ represents a face-on
orientation, and $C$ is a value between 0 and 1 and proportional to the bar's
strength.  The position angle of the bar, relative to the disc's apparent
minor axis, is denoted by $PA$.
For the reason discussed above, 
the above correction is minimal when the bar is parallel to the
projection axes (either major or minor axis). 
%
In passing we note the complication that as a disc's inclination increases
from face-on (0 degrees) to edge-on (90 degrees), the observed position angle
of the bar can {\it appear} increasingly aligned with the (apparent) position
angle of the disc (e.g.\ Debattista et al.\ 2002).  Measured disc inclinations
themselves can also be subject to the thickness of the disc, departures from
circularity (e.g.\ Andersen et al.\ 2001) and potential ellipticity gradients
in which the outer isophote shape depends on exposure depth.

\begin{figure}
\includegraphics[angle=270,scale=0.56]{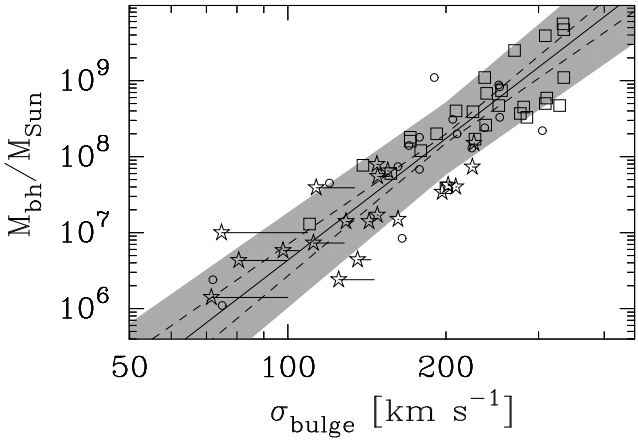}
\caption{
Adaptation of Figure~\ref{Fig5} such that (i) the central velocity
dispersions of the barred galaxies,  $\sigma_{\rm obs}$, 
have been adjusted to a new ``$\sigma_{\rm bulge}$'' value 
following equation~\ref{eq_Fran} using $C=0.7$, and (ii) 
the lines and shaded area now trace the $M_{\rm bh}$-$\sigma$ relation defined
by the non-barred galaxies, as given by the second last entry in Table~\ref{Tab2}. 
Squares, stars and circles denote elliptical, barred and non-barred disc 
galaxies, respectively.  The small lines emanating from the barred galaxies shows
their previous location in Figure~\ref{Fig5}.
}
\label{Fig7}
\end{figure}

In an effort to test if bar dynamics may be a viable (partial) explanation for
the offset barred galaxies in the $M_{\rm bh}$-$\sigma$ diagram, we have, for
an initial rough investigation, taken the disc inclination, position angle and
$V_{\rm rot}$ measurements from Hyperleda (see Table~\ref{Tab4}), and assumed
a $C$-value of 0.7 for all.  We note that the actual $C$-value is currently
unknown, and likely to be different for each galaxy, and we caution that the
heterogeneous nature of the data that goes into HyperLeda will also be a
source of scatter.  Given the simplistic nature of equation~\ref{eq_Fran}, we
were content, as a first approach, to estimate the bar's position angle by
eye.  Figure~\ref{Fig7} shows how the barred galaxies shift when using our
rough estimate of $\sigma_{\rm bulge}$ rather than $\sigma_{\rm obs}$.  One
can see that, as expected, the corrections are significant only in the cases
where $\sigma_{\rm obs}$, and thus M$_{BH}$, are small, and that it shifts
these cases towards smaller $\sigma_{\rm bulge}$ values.  

Depending on the values of $C$, this correction has the potential to improve
the correlation.  Obviously though, this work requires greater investigation
than we are able to afford here.  We also again remark that we have not dealt
with the second, independent phenomena known as sigma-drops, which may
potentially even account for some of the {\it apparent} over-shoot using
equation~\ref{eq_Fran}.

\begin{table}
\caption{Disc rotations and inclinations, and bar orientations}
\label{Tab4}
\begin{tabular}{@{}lcccc@{}}
\hline
Gal. Id.  & Type & $V_{\rm max}$ &  incl.$_{\rm disc}$  &  P.A.$_{\rm bar}$ \\
          &      & km s$^{-1}$   &  deg.                &  deg.  \\
\hline
IC 2560   & SBb  & 196   &  65.6   &  70 \\
Milky Way & SBbc & 220   &  90.0   &  20 \\
NGC  253  & SBc  & 194   &  78.0   &  45 \\ 
NGC 1023  & SB0  & 113   &  76.7   &  20 \\
NGC 1300  & SBbc & 167   &  49.3   &  60 \\
NGC 1316  & SB0  & 200   &  67.4   &  90 \\
NGC 2549  & SB0  &  70   &  90.0   &  20 \\
NGC 2778  & SB0  &  85   &  62.4   &  90 \\
NGC 2787  & SB0  & 182   &  66.2   &  10 \\
NGC 3079  & SBcd & 210   &  82.5   &  30 \\
NGC 3227  & SB   & 130   &  68.4   &  70 \\
NGC 3368  & SBab & 194   &  54.7   &  55 \\
NGC 3384  & SB0  & 167   &  90.0   &  00 \\
NGC 3393  & SBab & 158   &  31.0   &  10 \\
NGC 3489  & SB0  & 157   &  63.7   &  70 \\
NGC 4151  & SBab & 144   &  60.0$^*$   &  10 \\
NGC 4258  & SBbc & 208   &  72.0   &  60 \\
NGC 4596  & SB0  & 155   &  36.7   &  15 \\
NGC 4945  & SBcd & 167   &  90.0   &  45 \\
NGC 7582  & SBab & 195   &  68.2   &  20 \\
\hline
\end{tabular}

The maximum disc rotation velocities, $V_{\rm max}$, have been taken from
HyperLeda. 
The disc inclinations have been obtained from available photometry, under the
assumption that any intrinsic disc ellipticity is small and that the observed 
minor-to-major axis ratio is due to disc inclination 
($^*$ taken from Graham \& Li 2009). 
The rough position angle of the bars, P.A.$_{\rm bar}$, relative to the minor axis,
has been estimated by eye. 
\end{table}


%
%



\subsection{Nuclear star clusters}\label{Sec_NC} 

The offset nature, or rather the location below the upper envelope of
points, of some of the apparently non-barred AGN in the $M_{\rm bh}$-$\sigma$
diagram (Figure~\ref{Fig3}c) prompts one to consider alternative scenarios 
to bar dynamics.  A couple of non-barred galaxies with direct
SMBH mass measurements also appear to be offset from the barless $M_{\rm 
  bh}$-$\sigma$ relation (see Figure~\ref{Fig2}). 
If they are not pseudo bulges, in which the bar has disappeared, 
then something other than elevated
velocity dispersions may be responsible for their deviant nature. 

It has already been noted by Graham (2008a) that 
the inactive galaxy NGC~2778 has only a weak bar, 
yet this nucleated galaxy is 
below the barless $M_{\rm bh}$-$\sigma$ relation by 0.7 dex.
Curiously, the mass of the nuclear cluster in this galaxy is $\sim$5 times
(i.e.\ 0.7 dex) greater than 
the mass of its black hole.  Furthermore, 
the Milky Way and M32 have nuclear star cluster masses, $M_{\rm nc}$, 
which are 10$\times$ more
massive than their SMBH masses (Graham \& Spitler 2009, and references therein).   


Nuclear star clusters (Ferrarese et al.\ 2006; Wehner \& Harris 2006)
are now known to coexist with SMBHs at the centres of low- and
intermediate-mass spheroids (Filippenko \& Ho 2003; Graham \& Driver 2007a;
Gonz\'alez Delgado et al.\ 2008; Seth et al.\ 2008; Graham \& Spitler 2009).
The star clusters undoubtedly contribute, at some level, to the feeding of the
central black hole, whether by direct stellar infall (e.g., Lightman \&
Shapiro 1978; Merritt \& Vasiliev 2010) or through stellar winds (e.g., Ciotti
et al.\ 
1991; Soria et al.\ 2006; Hueyotl-Zahuantitla et al.\ 2010).  Schartmann et
al.\ (2010), for example, describe how stellar mass loss from the nuclear star
cluster (or nuclear disc, Davies et al.\ 2007) in NGC~1068 can feed the black
hole of this active Seyfert galaxy.  In unrelated work, Bekki \& Graham (2010)
argue why a seed SMBH in nuclear star clusters may be a necessary ingredient
to explain the absence --- through SMBH binary heating and ultimately erosion
--- of nuclear star clusters in massive galaxies built via hierarchical
merging.  Given the growing number of probable connections between SMBHs and
nuclear star clusters, which is no doubt yet to be fully appreciated, it seems
pertinent to explore the inclusion of the nuclear star cluster.

While we do not have nuclear star cluster masses for the AGN sample with
(reverberation-mapping)-derived SMBH masses, and can not therefore re-derive the $f$-factor, 
progress has commenced on acquiring star cluster masses for the local sample
of predominantly inactive galaxies with direct SMBH mass measurements (Graham
\& Spitler 2009).  In addition, Graham (2010, in prep.) tabulates yet more
galaxies having {\it both} a massive black hole and a nuclear star cluster.  
Following Graham \& Guzm\'an (2003) and Balcells et al. (2003, 2007), the
nuclear star cluster fluxes have been derived by simultaneulsy fitting for the
cluster, the host bulge and, when present, nuclear and large scale discs.
Failing to account for these components can bias the S\'ersic model used to
describe the bulge light, and thus bias the flux of the nuclear star cluster.
Foreground stars, background galaxies, and any apparent internal dust is
masked out before the seeing convolved models are fit to each galaxy's
distribution of stellar light.  The nuclear cluster fluxes are then converted
into masses following the methodology in Graham \& Spitler (2009).

\begin{figure}
\includegraphics[angle=270,scale=0.56]{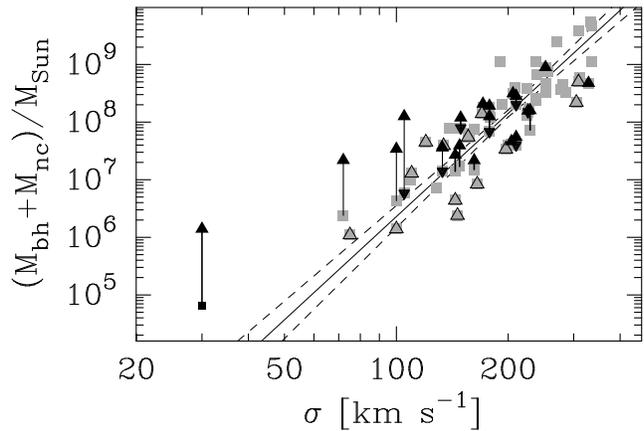}
\caption{
Adaptation of Figure~\ref{Fig5} using the 64 galaxies with direct SMBH masses
(plus NGC~4395, lower left).  The solid upward pointing arrow heads show the
location of the combined black hole masses (Graham 2008b and Table~\ref{Tab1}) plus the nuclear 
star cluster mass (Graham \& Spitler 2009 and Graham 2010, in prep.). 
For seven systems with only an upper limit to the nuclear cluster mass, a 
bi-directional arrow denotes the possible ($M_{\rm bh}+M_{\rm nc}$) mass range.  Galaxies 
with dust obscured nuclei, and/or significant AGN whose flux may hide a nuclear
star cluster, are plotted with open triangles on top of the mass of the SMBH. 
}
\label{Fig8}
\end{figure}

With this expanded data set, Figure~\ref{Fig8} explores the 
speculation by Graham \& Spitler (2009) that an insightful quantity to
plot on the vertical axis may be the combination of the SMBH {\it plus} the
nuclear star cluster mass.  Nayakshin et al.\ (2009) discuss how
competing feedback from a SMBH and a nuclear star cluster may explain
either end of such an ($M_{\rm bh}+M_{\rm nc}$)-$\sigma$ relation (see
also McLaughlin, King \& Nayakshin 2006 and 
Cantalupo 2010).  When both types of nuclei exist, it may make
sense to combine their masses, and Figure~\ref{Fig8} presents the first ever 
``($M$$+$$M$)-$\sigma$'' diagram showing this.  
Rather than treating galaxies as if they only have one type of nuclear
component, Figure~\ref{Fig8} shows that the low mass end of the $M_{\rm
bh}$-$\sigma$ diagram flattens when transformed into the ($M_{\rm bh}+M_{\rm
nc}$)-$\sigma$ diagram.  While too far off topic for the present paper, Graham
(2010, in prep.) expands this analyis to include (the transition to) low-mass galaxies
for which only the nuclear star cluster mass is known.

If NCs have a different velocity dispersion than their host bulge,
as is the case for NGC~205 (Carter \& Sadler 1990, see their figure~2), 
then they could bias measurements of the host bulge's central velocity
dispersion $\sigma$. 
However, for most galaxies the NC-to-(host bulge) flux ratio will
be small in typical ground-based aperture measurements, 
which could explain why observed ``sigma-drops'' are relatively rare, which is
not to say that they do actually exist in great numbers. 
High spatial resolution spectra would be desirable for pursuing this issue
of contamination at distances of the Virgo and Fornax galaxy clusters.
 
It is interesting that the galaxies with the smallest SMBHs in the
left hand side of Figure~\ref{Fig8} tend to be dusty Seyfert 2 galaxies with
AGN flux that contaminates the nuclear stellar flux, and thus any NC 
mass measurement.  These AGN with direct SMBH mass measurements are,
like the reverberation-mapped AGN, offset below (or rightward of) the
upper envelope of points in the $M_{\rm bh}$-$\sigma$ diagram.  Their 
AGN flux, and dusty central region,  may hide a nuclear star cluster and 
effectively bias these points low in the ($M_{\rm bh}+M_{\rm nc}$)-$\sigma$ diagram. 
Nonetheless, from Figure~\ref{Fig8} it is apparent that the slope at the low-mass end of the
($M_{\rm bh}$$+$$M_{\rm nc}$)-$\sigma$ diagram is reduced relative to the
$M_{\rm bh}$-$\sigma$
relation.  As discussed in Graham (2010, in prep.), this behavior appears 
to track the change in slope seen in the luminosity-(velocity dispersion)
diagram (e.g.\ De Rijcke et al.\ 2005; Matkovi\'c \& Guzm\'an 2005, and 
references therein), and is thus connected with the 
($M_{\rm bh}+M_{\rm nc}$)/$L_{\rm bulge}$ ratio. 

\subsection{Radiation Pressure} \label{Sec_RP}

Marconi et al.\ (2008) explored the effect that radiation
pressure (e.g.\ Mathews 1993), from the AGN's continuum emission, may
be having on the surrounding gas clouds, and thus the effect on measurements of the
virial product.  Such an outward force would counteract, at some
level, the inward force of gravity on the gas. Marconi et al.\ (2008) 
therefore argued that neglecting the radiation pressure in
reverberation mapping analyses results in a systematic
under-estimation of true black hole masses.  Armed
with flux measurements at 5100\AA, they empirically derived an
$f$-factor of 3.1$^{+1.3}_{-1.5}$, smaller than the value of 5.5$\pm$1.7
in Onken et al.\ (2004), but now with an extra luminosity-dependent
term.

Marconi et al.'s (2008) radiation pressure term will result in an upward
revision to the masses of high-accretion AGNs.  Given that Narrow Line
Seyfert 1 (NLS1) galaxies tend to have high accretion rates, and may reside
below the M$_{\rm bh}-\sigma$ relation (Wandel 2002; Mathur \& Grupe
2005; Wu 2009; but see Botte et al. 2005; Komossa \& Xu 2007; 
Decarli et al.\ 2008a), 
Marconi et al.\ (2008) point out that the apparent offset of the NLS1 galaxies
relative to the broad-line AGN may be due to past under-estimates of their
black hole masses. 
%

%

While Netzer (2009) concluded that radiation pressure is not
significant, and that the empirically-determined value by Marconi et
al.\ (2008) is too high, the rebuttal by Marconi et al.\ (2009)
counters this. What remains unclear is whether the slight reduction in
the scatter for the AGN $M_{\rm bh}-\sigma$ relation is due to an
improvement in the physical model or simply the addition of an extra
fitting parameter (see Peterson 2010).
%
Here we simply remark that a corrective term for radiation pressure would
act to drive the $f$-factor to even lower values.  That is, while our present
work has halved the commonly used value, it may still need to be revised
further downward. 


\subsection{Implications for AGN}\label{Sec_imp}


As a measure of the relationship between emission line widths and the
gravitational potential of the central SMBH, the $f$-factor is a direct
constraint on the kinematics and geometry of the broad-line region. An $f$
value of 3 is implied for the simplest model of a BLR (Netzer 1990). The next
level of sophistication in BLR modeling involves a flattening of the spatial
distribution. Thin discs, thickened discs, or composites of planar and random
motions have all been modeled by various groups (see discussions by Krolik
2001; Collin et al.\ 2006; Labita et al.\ 2006; Decarli et al.\ 2008b). The
uniqueness of any of these solutions remains unclear, but improved estimation
of both $f$ and the scatter around $f$ (to be expected with an
inclination-dependent BLR model) are positive steps towards understanding the
BLR geometry.


One consequence of our downward revision in the mean value of $f$ is
to reduce the black hole masses for any AGNs calibrated to the Onken
et al.\ (2004) result. Perhaps frustratingly, our lower value of $f$
gives masses quite similar to those which would have been produced by
the simplistic BLR model which had been in use prior to the work by
Onken et al.\ (modulo the relationship between FWHM and $\sigma_{\rm
line}$ that one adopts). Obviously, the change in $f$ directly impacts
any reverberation results that adopted the Onken et al.\ (2004) value
(Peterson et al.\ 2004, 2005; Metzroth et al.\ 2006; Bentz et al.\
2006, 2007, 2009a; Denney et al.\ 2006, 2009; Sergeev et al.\
2007; Grier et al.\ 2008).

Furthermore, the single-epoch methods of SMBH mass estimation which
were calibrated to the reverberation mapping results would be
similarly affected (Vestergaard \& Peterson 2006; McGill et al.\ 2008;
Wang et al.\ 2009; Greene, Peng \& Ludwig 2010). Because these single-epoch
methods are the primary basis for estimating black hole masses from
AGN spectra, the effects of changing $f$ can be far-ranging, from
individual objects to large AGN surveys (Barth et al.\ 2005; Woo et
al.\ 2006, 2008; Dong et al.\ 2007; Zhang et al.\ 2007; Treu et al.\
2007; Vestergaard et al.\ 2008; Shen et al.\ 2008a,b; 
Trump et al.\ 2009; Merloni et al.\ 2010; Greene et al.\ 2010;
Morgan et al.\ 2010; Lamastra et al.\ 2010).
%

For some applications, a new $f$ value has no significant ramifications: e.g.,
the evolution in the relationship between black holes and spheroids {\it in
AGN samples} 
(e.g. Peng et al.\ 2006a,b; Bennert et al.\ 2010; Decarli et al.\ 2010a,b); 
or the slope of the active 
BH mass function (e.g.\ Kelly, Vestergaard \& Fan 2009; Vestergaard \& Osmer
2009). In other cases, the fact that the modifications at present are uniform
to all BH masses means that the results are simply shifted by a factor of
$\sim 2$ (e.g., the Eddington ratio distributions of AGNs: Kollmeier et al.\
2006; the position of the peak in the black hole mass function: Kelly et al.\
2009).  In some circumstances, the shift in black hole masses may have a still
larger impact (e.g., in modeling the growth of black holes over cosmic time:
Yu \& Lu 2008; Vestergaard \& Osmer 2009).

However, it bears noting that, if $f$ differs for barred and
non-barred AGNs, changes in the bar fraction with redshift and/or with
galaxy mass could dramatically alter the conclusions of studies
unaffected by a simple $f$ shift (i.e., the slope of the BH mass
function, and the evolution of $M_{\rm BH}$-galaxy relationships).

As a final example of potential implications, we discuss the growing
evidence that massive black holes in high-redshift AGNs may pose a
challenge to the notion that such objects can grow from stellar-mass
seeds via normal (Eddington-limited) accretion in the time available
since the Big Bang (Dietrich \& Hamann 2004).  To explore the impact
of a revised $f$ value on such analyses, we focus on two particular
objects.  First, the $3\times10^{9}$~M$_{\odot}$ black hole found in
the $z=6.41$ AGN, SDSS J1148+5251 (Willott, McLure \& Jarvis 2003), was measured
by a mass equation which assumed $f=1$ (for a line width measured by
the FWHM). Our $f$ value would increase the BH mass for this object,
making it even harder to grow such a massive BH at such an early
time. In contrast, the BH mass in the $z=2.131$ AGN, Q0019+0107, of
9.5$\times10^{9}$~M$_{\odot}$ (Dietrich et al.\ 2009) would be halved,
and would therefore no longer require a seed black hole quite so
massive as 10$^{5}$~M$_{\odot}$. These two cases demonstrate the
complexity involved in comparing any SMBH masses in AGNs from the
literature, especially since the pedigrees of such estimates are
sometimes not clearly documented.

\section{Conclusions}

Using a sample of 64 galaxies with directly measured supermassive black hole
masses, the ``classical'' or ``standard'' $M_{\rm bh}$-$\sigma$ relation
(which contains galaxies of all morphological type) is shown here to have a
steeper slope than previously recognised.  Due to (i) the inclusion of more barred
galaxies, which tend to be offset from the non-barred galaxies (Graham
2008a,b) and had previously been under-represented in the $M_{\rm
bh}$-$\sigma$ diagram, along with (ii) increased black hole masses at the high-mass
end of the $M_{\rm bh}$-$\sigma$ diagram (e.g.\ Gebhardt \& Thomas 2009; Shen
\& Gebhardt 2009; van den Bosch \& de Zeeuw 2010), and (iii) the use of a 10 per
cent, rather than a rather optimistic 5 per cent, uncertainty on the velocity
dispersion, the slope is now 
somewhere between 5 and 6 depending on the type of regression used (see
Table~\ref{Tab2}).  
From a non-symmetrical regression of $\log M_{\rm bh}$ on
$\log \sigma$, and using a 10 per cent uncertainty on the velocity dispersion,
one has $\log(M_{\rm bh}/M_{\odot}) = (8.13\pm0.05) + (5.13\pm0.34)\log
[\sigma/200\, {\rm km\, s}^{-1}]$, with an r.m.s.\ scatter of 0.43 dex.
While, formally, the intrinsic scatter is 0.32 dex, we note that a more
detailed analaysis of the uncertainty in the velocity dispersions would be
welcome. 

If intermediate mass black holes exist (e.g.\ Greene \& Ho 2004, 2007), 
and an observational
selection bias is artificially truncating data in the $M_{\rm bh}$-$\sigma$
diagram below $M_{\rm bh} = 10^6 M_{\odot}$, then the above relation is biased. 
Using a regression of $\log \sigma$ on $\log M_{\rm bh}$ gives the bias-free
relation 
$\log(M_{\rm bh}/M_{\odot}) = (8.15\pm0.06) + (5.95\pm0.44)\log
[\sigma/200\, {\rm km\, s}^{-1}]$. 
When one doesn't know the morphological type of their galaxy, these
expressions are currently the best available estimators for $M_{\rm bh}$ from
measurements of $\sigma$.


As was first pointed out by Graham (2007a, 2008a,b) and Hu (2008), the
elliptical-only galaxies, and the non-barred galaxies, define tighter
relations with less scatter and a reduced slope than is obtained when using
the full galaxy sample.  Furthermore, the barred $M_{\rm bh}$-$\sigma$
relation is shown here to reside 0.47 and 0.45 dex below the elliptical-only
and barless $M_{\rm bh}$-$\sigma$ relations, respectively.  When one knows the
morphological type of their galaxy, these relations (given in
Table~\ref{Tab2}) are preferred for predicting $M_{\rm bh}$ from measurements
of $\sigma$ (as described in Section~\ref{Predict}).
The barless and elliptical-only $M_{\rm bh}$-$\sigma$ relations have a total 
r.m.s.\ scatter of 0.37 and 0.34 dex, respectively, 
and a slope of 4.95 and 4.87 when constructed with a symmetrical treatment of
the data.  A slope which is also consistent (at the 1-sigma level) with a value of
5 is obtained from a linear regression of $\log \sigma$ on $\log M_{\rm bh}$
(see Table~\ref{Tab2}).  

In terms of an integer slope for the $M_{\rm bh}$-$\sigma$ relation, when
constructed to address the ``Theorist’s Question'' (Novak et al.\ 2006), a
value of 5 is thus preferred for elliptical galaxies, in agreement with the
prediction by Silk \& Rees (1998).  Due to both new and refined SMBH masses,
and increased values at the high-mass end, this slope has increased from the
values around 4 that were reported by Graham (2008b), Hu (2008) and G\"ultekin
et al.\ (2009b).  The even steeper slope, close to a value of 6, which is
obtained for the full galaxy sample is due to the relatively offset nature of
the barred galaxies in the lower-left of the $M_{\rm bh}$-$\sigma$ diagram 
and the use of a linear regression not biased by sample selection which 
excludes BHs with $M_{\rm bh} < 10^6 M_{\odot}$.

Using a sample of 28 AGN with available virial products and host bulge
velocity dispersion measurements, we have explored their location in the 
$M_{\rm bh}$-$\sigma$ diagram. 
We have then derived a new $f$-factor for converting AGN virial
products (equation~\ref{Eq_f}) into black hole masses.  Our value
$f=2.8^{+0.7}_{-0.5}$ is a factor of two smaller than the commonly
used value $5.5\pm1.7$ (Onken et al.\ 2004, see also Woo et al.\ 2010
who reported $f=5.2\pm1.2$).  Moreover, this value might come down
even further due to processes that may be acting to enhance the offset
nature of AGN virial products from the $M_{\rm bh}$-$\sigma$ relation
defined by predominantly inactive galaxies with direct measurements of
their supermassive black hole.  For example, 
radiation pressure from the AGN may be
acting to partly counter-balance the inward force of gravity (on the
broad line region clouds) due to the black hole, possibly resulting in
anemic virial products (Marconi et al.\ 2008).  Correcting for this 
possibility will only lower the derived $f$-factor further. As such
we conclude that our reduced $f$-factor of $f=2.8^{+0.7}_{-0.5}$ 
($=2.6^{+0.6}_{-0.4}$ when excluding NGC~5548) may
still be an upper limit, and therefore some quasar and AGN masses in the 
literature are too high by at least a factor of 2.
Some of the implications of this are discussed in Section~\ref{Sec_imp}. 

This paper has also highlighted a number of issues.  First, 
the impact of galaxy type (i.e.\ elliptical, barred, non-barred) and thus sample selection 
on the $M_{\rm bh}$-$\sigma$ relation is 
important.  Depending on the relative numbers of barred and non-barred
galaxies, the best-fitting $M_{\rm bh}$-$\sigma$ relation changes.  
This of course undermines past 
efforts which had focussed on the exact value of the slope.  It also probably
voids several evolutionary studies based on earlier $M_{\rm bh}$-$\sigma$
relations. 
The present study is also not immune from this, although the construction of 
barred and non-barred $M_{\rm bh}$-$\sigma$ relations for galaxies with
directly measured SMBH masses is a positive step, plus this study is 
based on a larger and more representative sample of galaxies than ever before.
Second, as the black hole masses continue to be updated and refined,
systematic shifts in their masses can affect the $M_{\rm bh}$-$\sigma$
relation.  For example, 
van den Bosch \& de Zeeuw (2010) have shown how past assumptions that 
galaxies are either axisymmetric oblate spheroids or spherical, rather triaxial,
has resulted in an underestimation of their SMBH mass. 
Gebhardt \& Thomas (2009) have revealed how ignoring the dark matter's 
influence on the stellar dynamics can bias the stellar mass-to-light ratio resulting
in an underestimation of the SMBH mass. 
In the unlikely event that all of the current black hole
masses at the high mass end are doubled, i.e.\ increased by 0.3 dex, it would
result in a 10 per cent increase to the slope of the relation from $10^6$ to
$10^9 M_{\odot}$ (i.e.\ over a range of 3 dex in SMBH mass).
Third, this work confirms Graham's (2007a;2008a,b) and Hu's (2008) 
finding that the scatter about the $M_{\rm bh}$-$\sigma$
relation is quite large, and that velocity dispersion alone is therefore 
probably not the driving
force which dictates SMBH mass; if it was all that mattered, then the barred
galaxies would follow the same distribution as the non-barred galaxies in the
$M_{\rm bh}$-$\sigma$ diagram. 

There are several potential causes for this factor of $\sim$3 offset between
the barred and non-barred galaxies in the $M_{\rm bh}$-$\sigma$ diagram,
although it remains unclear what their various contributions are.
Possibilities include heightened velocity dispersion measurements due to (i)
elongated motions along a bar, (ii) vertical instabilities taking stars out of
the disc plane, (iii) rotational shear due to approaching and receding parts
of a disc within one's aperture, and iv) the variety of orbital shapes in and
around the bar.
The previously ignored contribution from massive nuclear star clusters
may also play a role (see Section~\ref{Sec_err}).  Nuclear star
clusters are also prevalent in low-mass, elliptical galaxies.
Including more of this galaxy type, to check for their offset behaviour in the 
$M_{\rm bh}$-$\sigma$ diagram, may 
be insightful. As a first step, we have presented the first ever $(M_{\rm
  bh}$$+$$M_{\rm nc})$-$\sigma$ diagram (Figure~\ref{Fig8}),
qualitatively showing how the slope at the low-mass end of the
$(M$$+$$M)$-$\sigma$ diagram is shallower than it is at the high-mass
end where nuclear star clusters are not detected.


\begin{table*}
\caption{AGN data}
\label{Tab3}
\begin{tabular}{@{}llll@{}}
\hline
\multicolumn{1}{l}{Gal.\ Id.}   &  Class.\       &  Virial Product          &  \multicolumn{1}{c}{$\sigma$}  \\
                &                &  [$10^6 M_{\odot}$]      &  \multicolumn{1}{c}{[km s$^{-1}$]} \\
\multicolumn{1}{l}{1}  &  \multicolumn{1}{l}{2}    & \multicolumn{1}{c}{3}    &  \multicolumn{1}{c}{4}   \\

\hline
\multicolumn{4}{c}{28 non-barred galaxies} \\
3C 120           & S0 (pec?)     & 10.1$^{+5.7}_{-4.1}$     &   162$\pm20^L$           \\  
Ark 120          & Sb pec        & 27.2$\pm$3.5             &   239$\pm36^M, 221\pm17^N$, {\boldmath $224\pm15$} \\  
Arp 151          & S0 pec        & 1.22$^{+0.17}_{-0.23}$$^A$ &   124$\pm12^O, 118\pm4^P$, {\boldmath $119\pm4$} \\
Mrk 110          & Sc            & 4.56$\pm$1.08            &   86$\pm13^Q, 91\pm25^N$, {\boldmath $87\pm12$} \\
Mrk 202          & Sc            & 0.26$^{+0.16}_{-0.11}$$^A$ &   86$\pm14^O, 78\pm3^P$, {\boldmath $78\pm3$}    \\ 
Mrk 279          & S0            & 6.35$\pm$1.66            &   197$\pm12^N$   \\
Mrk 290          & E1            & 4.42$^{+0.67}_{-0.67}$$^B$ &   ...         \\
Mrk 335          & S0/a          & 2.58$\pm$0.67            &   ...         \\ 
Mrk 509          & E2            & 26.1$\pm$2.1             &   ...         \\
Mrk 590          & SA(s)a        & 8.64$\pm$1.34            &   194$\pm20^M, 189\pm6^N, 169\pm28^L$, {\boldmath $189\pm6$} \\ 
Mrk 1310         & E             & 0.41$^{+0.16}_{-0.16}$$^A$ &   50$\pm16^O, 84\pm5^P$, {\boldmath $81\pm5$}    \\
NGC 4395         & Sm            & 0.065$\pm0.020$$^C$        &   25-35$^R$, {\boldmath $30\pm2.5$}  \\ 
NGC 4748         & Sa            & 0.47$^{+0.19}_{-0.23}$$^A$ &  105$\pm13^P$              \\
NGC 5548         & (R’)SA(s)0/a  & 12.1$^{+0.5}_{-0.5}$$^J$   &   183$\pm27^Q, 201\pm12^N, 195\pm13^P$, {\boldmath $197\pm8$}  \\
PG 0026+129      & E1            & 71.4$\pm$17.4            &   ...         \\
PG 0052+251      & Sb            & 67.3$\pm$13.7            &   ...         \\
PG 0804+761      & E3            & 126.4$\pm$14.5           &   ...         \\
PG 0844+349      & Sa            & 16.8$\pm$7.0             &   ...         \\
PG 0953+414      & E4            & 50.1$\pm$10.7            &   ...         \\
PG 1226+023      & E3            & 160.6$\pm$34.1           &   ...         \\  
PG 1307+085      & E2            & 80.1$\pm$22.1            &   ...         \\
PG 1411+442      & E4            & 80.5$\pm$26.5            &   ...         \\
PG 1426+015      & E2            & 236.7$\pm$69.8           &   217$\pm15^S$             \\  
PG 1613+658      & E2            & 50.7$^{+23.4}_{-23.5}$   &   ...                      \\  
PG 1617+175      & E2            & 108.0$\pm$25.3           &   183$\pm47^T$             \\  
PG 1700+518      & E1            & 142$^{+33}_{-30}$        &   ...                      \\
PG 2130+099      & (R)Sa         & 6.9$\pm2.7$$^D$            &   172$\pm46^T$ \\  
SBS 1116+583A    & Sc            & 1.05$^{+0.38}_{-0.34}$$^A$ &   50$\pm18^O, 92\pm4^P$, {\boldmath $90\pm4$}    \\
\multicolumn{4}{c}{14 barred galaxies}  \\
Fairall 9        & SBa           & 46.3$\pm$10.0            &   228$\pm20^U$                 \\
Mrk 79           & SBb           & 9.52$\pm$2.61            &   130$\pm20^Q, 130\pm12^N$, {\boldmath $130\pm10$}  \\
Mrk 142          & SB0/a         & 0.40$^{+0.14}_{-0.15}$$^A$ &  ...          \\
Mrk 817          & SBc           & 9.46$\pm1.24$$^E$          &   142$\pm21^Q, 120\pm15^N$, {\boldmath $127\pm12$}  \\
NGC 3227         & SAB(s) pec    & 1.39$^{+0.29}_{-0.31}$$^B$ &   131$\pm11^M, 136\pm4^N, 128\pm13^L$, {\boldmath $135\pm4$} \\
NGC 3516         & (R)SB(s)      & 5.76$^{+0.51}_{-0.76}$$^B$ &   144$\pm35^V, 164\pm35^V, 181\pm5^N$, {\boldmath $180\pm5$} \\
NGC 3783         & (R’)SB(r)a    & 5.41$\pm$0.99            &    95$\pm10^M$                 \\ 
NGC 4051         & SAB(rs)bc     & 0.31$^{+0.10}_{-0.09}$$^B$ &    84$\pm9^Q, 89\pm3^N, 88\pm13^L$, {\boldmath $88\pm3$} \\
NGC 4151         & (R’)SAB(rs)ab & 8.31$^{+1.04}_{-0.85}$$^F$ &    93$\pm14^Q, 97\pm3^N, 119\pm26^L$, {\boldmath $97\pm3$}    \\   
NGC 4253         & SBa           & 0.32$^{+0.28}_{-0.25}$$^A$ &    93$\pm32^P$                 \\       
NGC 4593         & (R)SB(rs)b    & 1.8$^{+0.4}_{-0.4}$$^G$    &   124$\pm29^L, 135\pm6^N$, {\boldmath $135\pm6$}   \\ 
NGC 6814         & SBbc          & 3.36$^{+0.63}_{-0.64}$$^A$ &   115$\pm18^L, 95\pm3^P$, {\boldmath $96\pm3$}     \\ 
NGC 7469         & (R’)SAB(rs)a  & 2.21$\pm$0.25            &   152$\pm16^M, 131\pm5^N$, {\boldmath $133\pm5$}     \\
PG 1229+204      & SBc           & 13.4$\pm$6.2             &   162$\pm32^T$                 \\  
\multicolumn{4}{c}{4 excluded galaxies}  \\
1E 0754.6+3928   &  ?            & 21.0$^{+40.3}_{-12.6}$$^H$ &   ...         \\
3C 390.3         & Sa            & 52.3$\pm$11.7            &   240$\pm36^W, 273\pm16^N$, {\boldmath $268\pm15$}  \\
IC 4329A         & SA0           & 1.80$^{+3.25}_{-2.16}$$^I$ &   122$\pm13^M$  \\
PG 1211+143      & E2            & 26.5$\pm7.9$$^{I,K}$       &   ...         \\
\hline
\end{tabular}
\noindent



Column 1. Galaxy Identification.
Column 2. Galaxy Classification from NED, except for Fairall~9 and PG~1229+204 which have been denoted as barred by Bentz et al.\ (2009b, their Table~5), plus Mrk 202 and SBS 1116+583A whose type has been assigned based on SDSS images and colour. 
Column 3. Virial Product (and also black hole mass if $f=1$) using the data from Peterson et al.\ (2004) unless otherwise noted.
Column 4. Stellar velocity dispersions; the bold value is the weighted mean.\\
References and notes: 
$^A$ Bentz et al.\ (2009a), $H\beta$ line. 
$^B$ Denney et al.\ (2010). 
$^C$ Peterson et al.\ (2005). 
$^D$ Grier et al.\ (2008).  
$^E$ Weighted mean from Peterson et al.\ (2004) and Denney et al.\ (2010). 
$^F$ Bentz et al.\ (2006).  
$^G$ Denney et al.\ (2006). 
$^H$ Sergeev et al.\ (2007). 
$^I$ Time-lag uncertain (Peterson et al.\ 2004) and thus Virial Product not used. 
$^J$ Weighted mean from Peterson et al.\ (2004), Bentz et al.\ (2009a) and Denney et al.\ (2010).
$^K$ H$_{\beta}$ and H$_{\gamma}$, but not H$_{\alpha}$, line used here. 
$^L$ Nelson \& Whittle (1995).
$^M$ Onken et al.\ (2004).
$^N$ Nelson et al.\ (2004).
$^O$ Greene \& Ho (2006).  
$^P$ Woo et al.\ (2010). 
$^Q$ Ferrarese et al.\ (2001), using their suggested 15 per cent uncertainty.
$^R$ Filippenko \& Ho (2003, their section~3). 
$^S$ Watson et al.\ (2008). 
$^T$ Dasyra et al.\ (2007). 
$^U$ Oliva et al.\ (1995).  
$^V$ Arribas et al.\ (1997).
$^W$ Green et al.\ (2004).
\end{table*}

\section{acknowledgment}
We are grateful for Monica Valluri's comments on an early version of this
paper, in particular her words of caution regarding the accuracy and
compatibility of literature velocity dispersions. 
E.Athanassoula acknowledges partial support from grant ANR-06-BLAN-0172. 
We acknowledge use of the HyperLeda database (http://leda.univ-lyon1.fr).
This research has made use of the NASA/IPAC Extragalactic Database (NED).  

After completion of this work we became aware of four additional velocity
dispersion measurements for our AGN sample.  Using a direct filtering method,
Garcia-Rissmann et al.\ (2005) report velocities of 116$\pm$20, 153$\pm$24,
83$\pm$11 and 125$\pm$12 km s$^{-1}$ for NGC~3783, 4593, 6814 and 7469,
respectively.  Within the quoted errors, these values agree with those used in
this paper.

\label{lastpage}
\end{document}